\documentclass{article}

\usepackage{arxiv}

\usepackage[utf8]{inputenc} 
\usepackage[T1]{fontenc}    
\usepackage{hyperref}       
\usepackage{url}            
\usepackage{booktabs}       
\usepackage{amsfonts}       
\usepackage{nicefrac}       
\usepackage{microtype}      
\usepackage{lipsum}

\usepackage{graphicx}
\usepackage{amsmath,amssymb,amsfonts, epsf}
\usepackage{algorithmic}
\usepackage[ruled,vlined]{algorithm2e} 
\usepackage{breqn}
\graphicspath{{images/}}
\usepackage{nomencl}
\usepackage{xcolor}

\title{Hybrid deep fault detection and isolation: Combining deep neural networks and system performance models}

\author{
  Manuel Arias Chao \\
  Chair of Intelligent Maintenance Systems\\
  ETH Zurich, Switzerland\\
  \texttt{manuel.arias@ethz.ch} \\
   \And
  Chetan Kulkarni \\
  SGT, Inc.\\
  NASA Ames Research Center, USA\\
  \texttt{chetan.s.kulkarni@nasa.gov} \\
   \And
  Kai Goebel \\
   Operation and Maintenance Engineering\\
  Lule{\aa} University of Technology, Sweden\\
  \texttt{kai.goebel@ltu.se} \\
   \And
  Olga Fink \\
  Chair of Intelligent Maintenance Systems\\
  ETH Zurich, Switzerland\\
  \texttt{ofink@ethz.ch} \\
}

\begin{document}
\maketitle

\begin{abstract}
With the increased availability of condition monitoring data on the one hand and the increased complexity of explicit system physics-based models on the other hand, the application of data-driven approaches for fault detection and isolation has recently grown. While detection accuracy of such approaches is generally very good, their performance on fault isolation often suffers from the fact that fault conditions affect a large portion of the measured signals thereby masking the fault source. To overcome this limitation, we propose a hybrid approach combining physical performance models with deep learning algorithms. Unobserved process variables are inferred with a physics-based performance model to enhance the input space of a data-driven diagnostics model. The resulting increased input space gains representation power enabling more accurate fault detection and isolation.

To validate the effectiveness of the proposed method, we generate a condition monitoring dataset of an advanced gas turbine during flight conditions under healthy and four faulty operative conditions based on the Aero-Propulsion System Simulation (C-MAPSS) dynamical model. We evaluate the performance of the proposed hybrid methodology in combination with two different deep learning algorithms: deep feed forward neural networks and Variational Autoencoders, both of which demonstrate a significant improvement when applied within the hybrid fault detection and diagnostics framework. 

The proposed method is able to outperform pure data-driven solutions, particularly for systems with a high variability of operating conditions. It provides superior results both for fault detection as well as for fault isolation. For the fault isolation task, it overcomes the smearing effect that is commonly observed in pure data-driven approaches and enables a precise isolation of the affected signal. We also demonstrate that deep learning algorithms  provide a better performance on the fault detection task compared to the traditional machine learning algorithms.
\end{abstract}

\keywords{Hybrid Diagnostics \and Deep Learning \and C-MAPSS}

\section{Introduction} \label{sec:Introdution}

Increasing amounts of condition monitoring (CM) data from complex engineered systems, both in terms of the number of sensors as well as in terms of the sampling frequency, and advancements in machine and deep learning algorithms provide an untapped potential to extract information on asset health condition. Concretely, deep learning algorithms have demonstrated an excellent ability to learn the system behaviour directly from large volumes and variety of the condition monitoring signals and therefore decreased the need of manual feature engineering. As a result, deep learning-based solutions have been increasingly applied to complex learning tasks in prognostics and system health management (PHM) of complex systems \cite{Zhao2016, Khan2018}.

Since machine and deep learning algorithm rely on learning patterns from representative examples, one of the major challenges in applying deep learning algorithm for fault detection and diagnostics tasks is the lack of labeled data, i.e. a lack of a sufficient number of representative samples of known fault patterns. Only a representative dataset of possible fault types would enable the algorithms to learn all the characteristic patterns of the specific faults and provide very good fault detection and isolation capabilities. Because faults are rare in complex safety critical systems, such as aviation propulsion systems, it is unfeasible to obtain sufficient samples from all possible fault types that can potentially occur. However, most of the previous research in fault detection and diagnostics has been focusing on defining the problem of fault detection and diagnosis as a classification task and rather tackling the problem of imbalanced datasets for the faulty classes \cite{wang2013online, xu2007power, zhang2018imbalanced}.  

In this paper, we consider the case where we have only information on the healthy class, the number and nature of the fault classes are, however, not known in advance. This is a more realistic task for the practical applications but also a more difficult task compared to the case where the available labeled data samples already cover the essential information on the number and type of classes and the new observation only fall in the category of already known classes. As an additional degree of difficulty, we focus particularly on systems that are operated under varying conditions that are frequently changing. An example for such systems are aircraft engines experiencing continuous changes (transients) on the flight conditions. 

One of the previously most common approaches for the case when only healthy system conditions are available for model development is based on signal reconstruction and the subsequent analysis of the residuals between the monitored and reconstructed signals \cite{baraldi2015robust, hu2017fault}. Robust decision boundaries are crucial in this case for the performance of the algorithms. If condition monitoring signals are highly correlated, a so called smearing effect can occur influencing not only signals directly affected by the fault but also causing deviations in correlated signals that do not contain any information on the fault. This effect makes it difficult or even impossible to isolate the root cause of the fault if the fault isolation is solely based on the residuals \cite{hu2017fault}.

Recently, a new integrated fault diagnosis approach was introduced, combining feature learning with a one-class classification for the fault detection and a subsequent analysis of the residuals for the fault isolation task \cite{michau2017deep}. This solution strategy aims to map the observed healthy operation to a healthy class and later discriminate if the operating condition of interest with unknown health state follows the learned pattern of the healthy system conditions. The detection accuracy of such approaches is generally very good when the available healthy operating conditions used for training are clearly representative of the conditions under analysis. 

Varying operating conditions create a shift in the underlying distributions of the CM data. Training an algorithm on the combined representation of these operating conditions with a limited number of samples may result in an unsatisfactory performance of the algorithms since the fault characteristics may be masked by the variability of operating conditions. 

If the operating conditions are too dissimilar, a possible way to address this challenge would be to develop dedicated algorithms for each of the operating conditions and switch between the different algorithms depending on the operating condition of the current observation. Another way to benefit from the experience of several operating conditions is to apply domain adaptation and align the underlying distributions in the feature space \cite{wang2019domain}, enabling thereby the transfer of the experience between the different operating conditions. However, such alignment requires at least some labels in the training dataset for the fault types which we don't have for the selected problem setup.  

In this work, we focus on the challenging problem of fault detection and diagnostics under varying operating conditions and highly correlated signals. We propose a framework and a method that combines physics-based models and deep learning algorithms and is particularly targeting the case when faulty samples are not available during model development.  

Complex systems can be modelled at various levels of detail, ranging from simple algebraic relations to full 3D-description of the process. In this range, thermodynamic models (a.k.a. 0-D models) of different levels of fidelity are generally available for design or control of complex systems. These models typically a moderate computational load and yet are able to predict process measurements (e.g., temperatures, pressures, air mass flow rates, rotational speeds) as well as global system performance (e.g. efficiencies and power). Furthermore, system performance models offer access to unmeasured variables that might be more sensitive to fault signatures and consequently can improve fault detection and isolation.

In the proposed framework, unobserved process variables are inferred with a physics-based performance model to enhance the input space of a data-driven diagnostics model. The resulting increased input space gains representation power enabling more accurate fault detection and isolation. 

The focus of the proposed method is on fault detection and isolation for complex industrial assets that are operated under varying conditions. The main benefit of the proposed method arises particularly for systems for which we don't have sufficient labels to develop classification algorithms and for which pure data-driven approaches with a single model combining data from all the operating conditions provide unsatisfactory performance for fault detection and isolation.

The proposed hybrid framework can be combined with any deep learning algorithm. To demonstrate this, we combine it with a feed-forward neural network, a Variational Autoencoder and a vanilla autoencoder. To validate the fault detection and isolation capability of the proposed method, we generate a new dataset of an turbofan engine during flight conditions under healthy and four faulty operative conditions. The dataset was synthetically generated with the Commercial Modular Aero-Propulsion System Simulation (C-MAPSS) dynamical model \cite{Frederick2007}. Real flight conditions as recorded on board of a commercial jet were taken as input to the C-MAPSS model \cite{DASHlink}.The evaluated case study comprises simulated flight conditions under healthy and four faulty operating conditions.

To assess the effectiveness of the proposed hybrid framework, we first evaluate different deep learning architectures and then compare the performance to 1) pure deep learning algorithms with the same architecture as those applied within the hybrid approach; 2) a standard machine learning algorithm, the one-class support vector machines (OC-SVM) \cite{Scholkopf2000}; 3) an alternative hybrid framework based on residuals between the performance model and the real observed condition monitoring data in combination with the real condition monitoring data as input.  

We demonstrate that the proposed framework is able to outperform pure data-driven solutions, particularly for systems with a high variability of operating conditions. It provides superior results both for fault detection as well as for fault isolation. For the fault isolation task, it overcomes the smearing effect that is commonly observed in pure data-driven solutions and enables a precise isolation of the affected signal. We also demonstrate that deep learning algorithms provide a better performance on the fault detection task. 






\section{Related work} \label{sec:Related}

Data-driven and physics-based approaches have their advantages and limitations when applied as stand-alone approaches. While physics-based approaches do not require large amounts of data and retain the interpretability of a model, they are generally limited by their high complexity or incompleteness. On the contrary, data-driven approaches are simple to implement and are able to discover complex patterns from large volumes of data but are limited by the representativeness of the training datasets. The combined use of data-driven and physics-based approaches has the potential to lead to performance gains by leveraging the advantages of each method. 

Different solutions have been proposed to combine physics-based models and data-driven algorithms. Depending on what type of information is processed and how the pieces of information are combined, different types of hybrid architectures can be created. In the following, some of the proposed architectures that are the most comparable to the proposed framework are presented and discussed. 
 
\begin{figure}[ht]
\centering
\includegraphics[width=8.5cm]{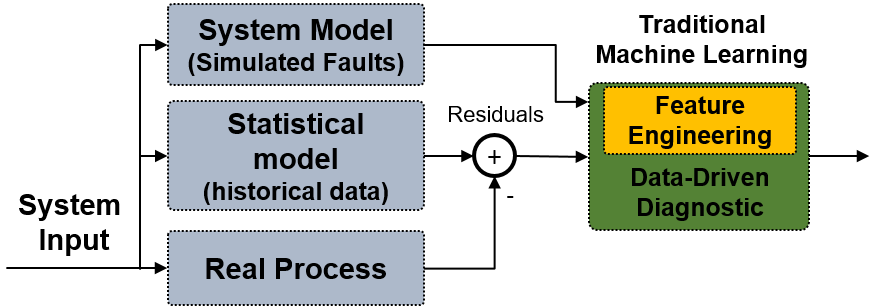}
\caption{Overall architecture of the hybrid diagnostics approach in \protect\cite{Frank2016}. The traditional machine learning algorithms take as input synthetic data of a healthy and faulty system that are generated with a system model and combined with the residuals between measurements and system performance estimation from a statistical model.}
\label{fig:hybrid_model_frank}
\end{figure}

Frank et al. \cite{Frank2016} explore the use of a hybrid approach where synthetic data of a healthy and faulty system are generated with a high-fidelity system model and used as input to traditional data-driven algorithms. Within the hybrid architecture, a range of traditional data-driven machine learning algorithms with an additional feature engineering step was explored, including random forests and support vector machines. The output of the system model is subsequently combined with residuals between measurements and system performance estimation from a statistical model (i.e. generated based on historical data). This architecture (see Figure \ref{fig:hybrid_model_frank}) is applied to diagnostics problems of abnormal energy consumption in buildings resulting from faulty equipment such as faults of air conditioners, chillers, dampers, and fan motors. 

\begin{figure}[ht]
\centering
\includegraphics[width=8.5cm]{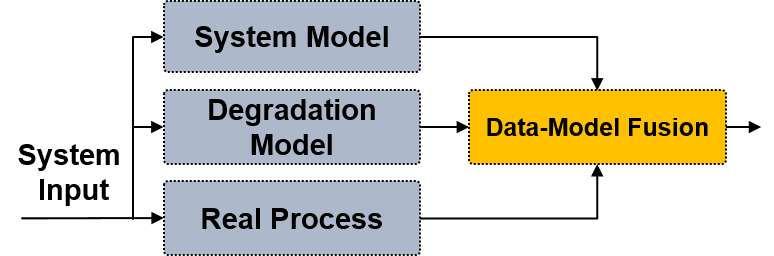}
\caption{Overall architecture of the parallel hybrid diagnostics framework in \protect\cite{Hanachi2017}. A particle filter is used as a fusion mechanism to aggregate the diagnostic results from measurement signals and degradation models.}
\label{fig:hybrid_model_hanachi}
\end{figure}

Hanachi et. al. \cite{Hanachi2017} use a parallel hybrid approach to diagnostics of gas turbines. In this approach, empirical (i.e. data-driven) fault transition models and physics-based system models perform the state assessment of the process at hand. Particle filter is used as a fusion mechanism to aggregate the diagnostic results from the measurement signals and degradation models (see Figure \ref{fig:hybrid_model_hanachi}).

\begin{figure}[ht]
\centering
\includegraphics[width=8.5cm]{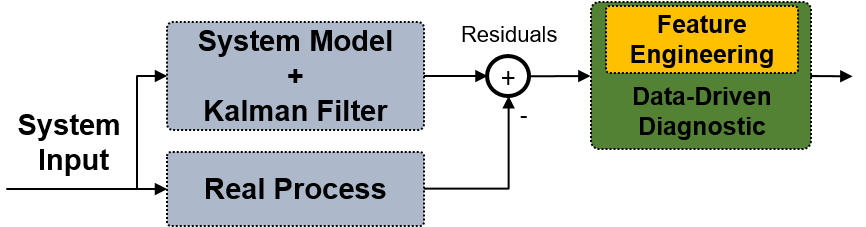}
\caption{Overall architecture of the residual-based hybrid diagnostics in \protect\cite{Rausch2005}. Feature engineering is carried out for the residuals between Kalman Filter estimates and sensor readings and are used as input to an SVM classifier.}
\label{fig:hybrid_res_rausch}
\end{figure}

A further possible architecture combining model-based and data-driven approaches uses first the system model to reason over the process and then a data-driven classifier that distinguishes between the different fault classes. Rausch et al. \cite{Rausch2005} use such an architecture for online fault diagnostics of turbofan engines (see Figure \ref{fig:hybrid_res_rausch}). In their approach, features engineered from the residuals between estimates of Extended Kalman Filter (EKF) and sensor readings are used as input to a machine learning classifier (SVM model).

\begin{figure}[ht]
\centering
\includegraphics[width=8.5cm]{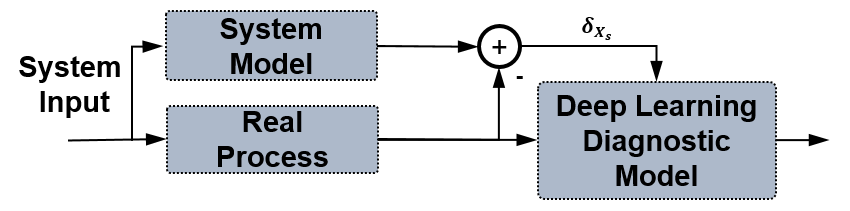}
\caption{Overall architecture of the residual-based diagnostics approach. The deep learning diagnostics model receives as input the system inputs (i.e. scenario-descriptor operating conditions) and the residual between sensor readings and estimated model responses $\delta_{x_s}$}
\label{fig:hybrid_res}
\end{figure}

Residual-based approaches are the standard for model-based diagnostics of aircraft engines. A generic residual-based diagnostics approach involves two major tasks. First, discrepancies between measurements and the expected healthy model responses are computed. In a second step, the residuals, that encode the potential the impact of degraded or faulty system behaviour, are processed with a fault detection and isolation (FDI) logic to create the diagnosis report \cite{Borguet2012}. Concretely, residuals can be fed as input to a deep-learning diagnostic algorithm in addition or instead of the measurements. Therefore, the fault detection and isolation logic is discovered by a deep neural network. Figure \ref{fig:hybrid_res} shows a block diagram of a residual-based hybrid diagnostics framework where deep learning diagnostics algorithm receives as input the scenario-descriptor operating conditions and the residual between sensor readings and estimated model responses.

Recently, several approaches of physics-guided machine learning have been proposed, where physical principles are used to inform the search of a physically meaningful and accurate machine learning model. The architecture proposed in \cite{Jia2018}, for example, enhances the input space to a data-driven system model with outputs from a physics-based system model. As a result, the dynamical behaviour of the system could be approximated more accurately.

\begin{figure}[ht]
\centering
\includegraphics[width=8cm]{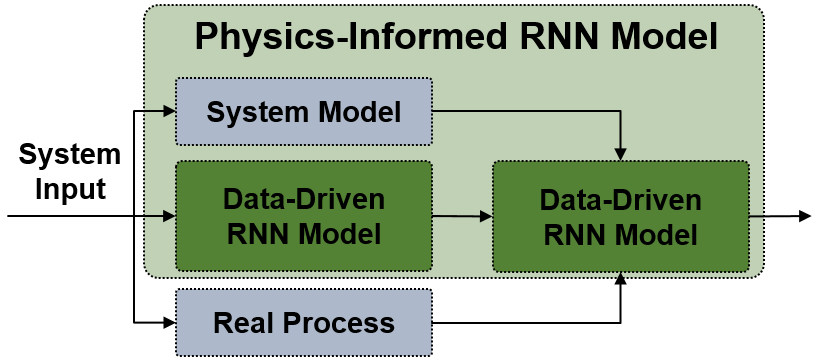}
\caption{Overall architecture of the physics-informed recurrent neural network in \protect\cite{Nascimento2019}}
\label{fig:hybrid_model_rai}
\end{figure}

In another variation of the physics-guided machine learning idea, a recurrent neural network cell was modified to incorporate the information from the system model at an internal state of the dynamical system (see Figure \ref{fig:hybrid_model_rai}). A related idea was applied to a variety of prognostics problems, such as in \cite{Nascimento2019, Dourado2019, Yucesan2019}. 

Contrary to previous hybrid architectures, the framework proposed in this paper leverages inferred \textbf{unobserved virtual sensors} and the \textbf{unobservable model parameters} to enhance the input space to a tailored deep learning-based FDI algorithm (see Figure \ref{fig:hybrid_cal}).

\section{Proposed Methodology} \label{sec:Methodology}

\subsection{Calibration-Based Hybrid Diagnostics} \label{sec:hydrid}
Physics-based performance models of different levels of fidelity are generally available for design or control of complex engineered systems. System performance models are represented mathematically as coupled systems of nonlinear equations. The inputs of the model are divided into scenario-descriptor operating conditions $w$ and model parameters $\theta$. The output of the system model is not limited to estimates of measured physical properties values $\hat{x}_s$ but also provides unmeasured properties $x_v$ (i.e. \textit{virtual sensors}). As there is no description given by an explicit formula, the nonlinear performance model is denoted as 
\begin{equation} \label{eq:model}
  [\hat{x}_s,x_v] = S(w,\theta)
\end{equation}
Performance models provide additional information that is not part of the condition monitoring signals and may be relevant for detecting faults. Therefore, we propose to make use of modelled variables $[\hat{x}_s, x_v, \theta]$ as input to the deep learning diagnostics algorithm. Hence, the resulting hybrid diagnostic approach combines information from physics-based models with CM data (i.e. $[w, x_s]$) and uses this enhanced input for subsequent fault detection and isolation with a deep learning algorithm.

\begin{figure}[ht]
\centering
\includegraphics[width=8.5cm]{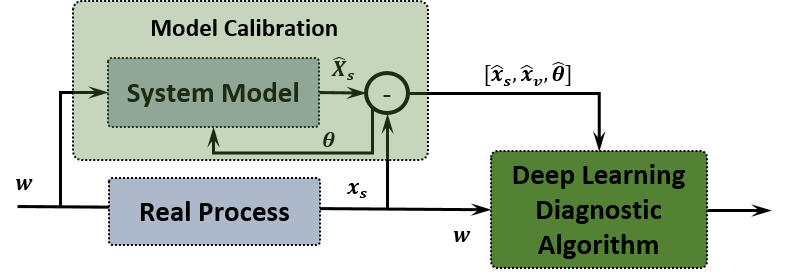}
\caption{Overall architecture of the calibration-based hybrid diagnostics framework. The deep learning diagnostics algorithm takes as input the scenario-descriptor operating conditions $w$, estimates of the condition monitoring signals ($\hat{x}_s$) and the virtual sensors ($\hat{x}_v$) and model parameters ($\theta$). }
\label{fig:hybrid_cal}
\end{figure}

However, to maximize the amount of relevant model information available for the generation of a data-driven diagnostics model, we propose to calibrate the system performance model $S(w,\theta)$. Model calibration involves inferring the values of the model parameters $\theta$ that make the system response to reproduce closely the observations $x_s$. Hence, the information about system degradation (and ideally the fault signature) is encoded within the estimated model correcting parameters $\hat{\theta}$. The calibrated model also provides high confidence estimates of process variables $\hat{x}_v$ that may be sensitive to fault signatures. Therefore, we propose to enhance the input space for the deep-learning diagnostic model with the process variables $[ \hat{x}_s, \hat{x}_v, \hat{\theta}]$ inferred with the system performance model. Figure \ref{fig:hybrid_cal} shows a block diagram of the proposed calibration-based hybrid diagnostic approach. The deep learning diagnostics model receives scenario-descriptor operating conditions $w$ and model variables $[\hat{x}_s, \hat{x}_v, \hat{\theta}]$ as input. The feedback arrow to the system model represents the calibration process for updating the model calibration parameters $\hat{\theta}$. Model calibration is a standard approach in several technical areas including traditional model-based diagnostics \cite{Brunell1999}, model-based control and performance analysis of system models \cite{AriasChao2015}.

The extended representation provided by the calibrated system model also provides additional interpretability and ability to isolate potential degradation root causes. The model parameters $\theta$ are indeed model tuning of the system components and hence a deteriorated behaviour of a sub component is precisely encoded in only one component of $\theta$ (i.e. $\theta_k$) while it is at the same time  manifested in the condition monitoring data and virtual sensors. As it will be shown in the case study (Section \ref{sec:Case_study}), this feature avoids the smearing characteristic of data-driven diagnostics models. An additional advantage of including the calibration processing step is that errors in the sensor readings can be detected and removed and therefore diagnostics process is more robust to sensor faults.

\begin{figure}[ht]
\centering
\includegraphics[width=8.5cm]{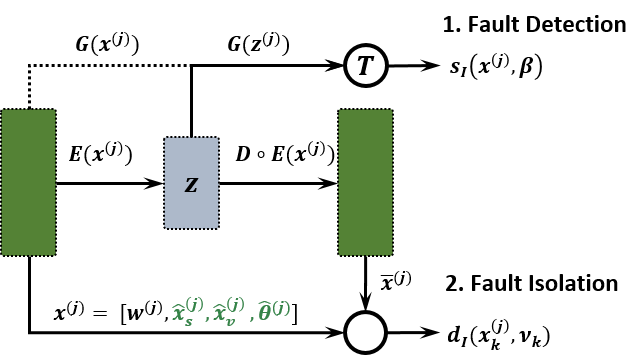}
\caption{Block diagram of the proposed fault detection and isolation algorithm within the proposes hybrid diagnostics framework. The FDI algorithm takes $x=[w,\hat{x}_s, \hat{x}_v, \theta]$ as input. A functional mapping $\mathcal{G}$ from the input ($x$) or an embedding representation of the input ($z$) to a target $\mathbf{T}$ is used to generate a similarity score $s_I(x^{(j)};\beta)$. Fault detection is performed based on $s_I(x^{(j)};\beta)$. The enhanced input signal is reconstructed ($\bar(x)$) with an autoencoder network. Fault isolation is performed based on the similarity score $d_I(x_k^{(j)};\nu_k)$. The variables in green are specific to the proposed hybrid approach.}
\label{fig:algorithm}
\end{figure}

In addition to the model calibration, a diagnostic report requires a clear fault detection and isolation algorithm, beyond the standard threshold-based logic. Therefore, we propose a tailored deep learning-based FDI algorithm shown in Figure \ref{fig:algorithm}. The proposed algorithm uses as input the extended representation provided by the calibrated system model ($x=[w,\hat{x}_s, \hat{x}_v, \theta]$) and computes a similarity score $s_I(x^{(j)};\beta)$. Fault detection is performed based on a clear logic on $s_I(x^{(j)};\beta)$. The enhanced input signal $x$ is reconstructed with an autoencoder network and fault isolation is performed based on the similarity score $d_I(x_k^{(j)};\nu_k)$. A detailed description of the proposed algorithm is covered in Section \ref{sec:methods}.

\section{Methods} \label{sec:methods}

\subsection{Problem Statement} \label{sec:Statement}
We aim at developing a diagnostic model able to detect and isolate fault types on complex systems operated under a large range of changing operating conditions. In our problem, we consider the situation where at model development time $t_a$, we have access to a dataset of condition monitoring signals and system model estimates of process variables. Certainty about healthy operative conditions are only known until a past point in time $t_b$ when an assessment of the system health was performed and declared healthy. Hence, at model development time, we only have access to the true healthy class for a portion of our data and fault signatures of an unknown number of fault types may be present in the remaining dataset. In particular, we consider the scenario where an evolving fault condition is actuality present but has not been detected due to the low intensity of the fault. In addition to the unlabelled data, an independent test set with increased levels of component degradation is provided. Our task is then to the detect the fault types in both, the unlabelled dataset and the test set. It should be noted that at $t_a$ an incomplete knowledge of the world is present. Hence, we have an \textit{open set} problem \cite{Scheirer2013} where we only know the initial healthy state but do not have any information on the faulty conditions. Therefore, not all possible classes are know at the model development stage and it is not even known how many fault classes may evolve. The formulation of the diagnostic problem addressed in this paper is formally introduced in the following. 

\begin{figure}[ht]
\centering
\includegraphics[width=8.6cm]{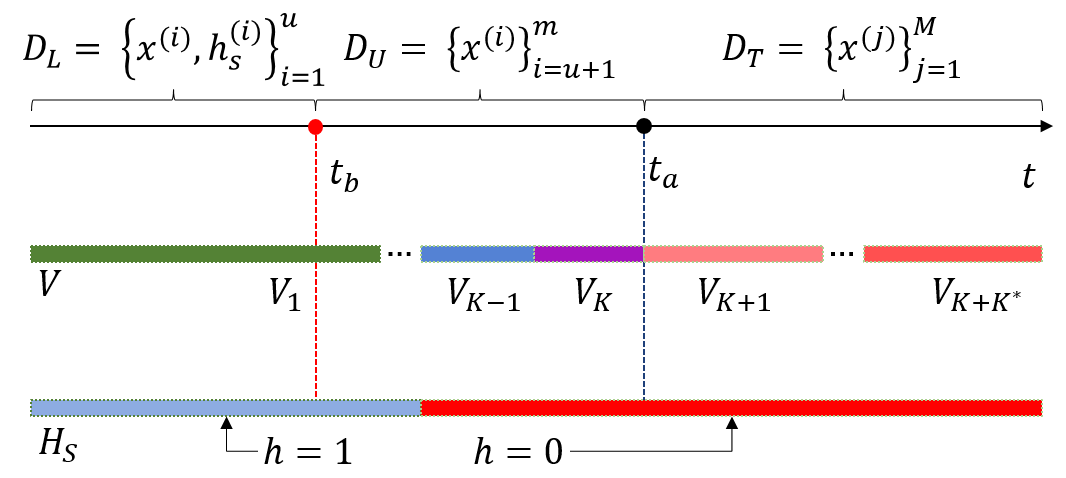}
\caption{Schematic representation of the problem. Training dataset $D$ has labelled ($D_L$) and unlabelled data ($D_U$). The test set ($D_T$) has only unlabelled samples. The true health condition any point in time is represented by the $H_S$ bar. Healthy condition are represented in blue and faulty in red. The true operative condition type within the data are represented by the $V$ bar. The healthy condition is shown in green; each fault type appear in a different color. $K$ fault classes are present in $D_U$ and $K_*$ in $D_T$.}
\label{fig:problem}
\end{figure}

Given is a multivariate time-series of condition monitoring sensors readings from one unit $X_s = [x_{s}^{(1)}, \dots, x_{s}^{(m)}]^T$, where each observation $x_{s}^{(i)} \in R^{p}$ is a vector of $p$ raw measurements taken at operative conditions $w^{(i)} \in R^{s}$. In addition we have available residuals between measurements and estimated healthy system responses (i.e. $\delta x_s^{(i)}$) and the output of a calibrated system model that provides inferred values of the model tuners $\theta^{(i)}$ and estimates of the sensors readings $\hat{x}_{s}^{(i)}$ and virtual sensors $\hat{x}_{v}^{(i)}$. Hence, in compact form, we denote the complete set of measured and inferred inputs as $X=\{({w}^{(i)}, \hat{x}_{s}^{(i)}, \hat{x}_{v}^{(i)}, \hat{\theta}^{(i)}, \delta_{m}^{(i)}))\}_{i=1}^{m}$. At model development time, the corresponding \textit{true} system's health state is partially known and denoted as $H_s = [h_s^{(1)}, \dots, h_s^{(m)}]^T$ with $h_s^{(i)} \in \{0,1\}$ where the healthy class is labeled as $h_s^{(i)}=1$. Therefore, our partial knowledge of the true health allows to define two subsets of the available data: a \textit{labeled} dataset $\mathcal{D}_{L} = \{(x^{(i)}, h_{s}^{(i)})\}_{i=1}^{u}$ with $h_s^{(i)}=1$ corresponding to known healthy operative conditions and an \emph{unlabeled sample} $\mathcal{D}_{U} = \{ x^{(i)})\}_{i=u+1}^{m}$ with unavailable health labels. In particular, we consider scenarios where $K$ unknown faults types are present in ${D}_{U}$. The fault types correspond to increasing intensities of the same fault mode (i.e. step-wise increases). The level of component degradation in $D_U$ is low (i.e. $\leq -1\%$ nominal conditions) and therefore we represent the situation where faults signatures are present but are not yet detected at analysis time. In addition, we test the generalization capability of our model to detect $K_{*}$ new faults of higher intensity in a test dataset $\mathcal{D_{T}} = \{(x_{*}^{(j)}\}_{j=1}^{M}$. An schematic representation of the problem is provided in Figure \ref{fig:problem}.

Given this set-up we first consider the problem of detecting the faulty operative within $\{{D}_{U}, {D}_{T}\}$ given only our healthy dataset $\mathcal{D}_{L}$ at time $t_a$. Hence our initial task is to estimate the health state (i.e $\mathbf{\hat{h}_s}$) on $\{{D}_{U}, {D}_{T}\}$. Furthermore, we aim to provide an isolation of the fault mode present.  We refer to $\textbf{V}=\{V_j|j=1, \dots, R\}$ as the partition of $\{{D}_{U}, \mathcal{D_{T}}\}$ according to the $R=K+K_{*}+1$ true but unknown fault types we aim to detect. For simplicity we will refer to the dataset $\{{D}_{U}, {D}_{T}\}$ as the combined test set that we denote as $D_{T+}$. 
\subsection{System Model Calibration} \label{sec:calibration-methods}
A conventional way to ensure that the system response follows observations $X_s$ is to infer the values of the model correcting parameters $\theta$ solving an inverse problem. Since both the measurement data and model parameters are uncertain, the process of estimating optimal correcting parameters is a stochastic calibration problem. Ideally, the calibration process aims at obtaining the posterior distribution of the calibration factors given the data $p(\theta|w, X_s)$. However, computing the whole distribution is generally computationally very expensive and therefore in most cases, point value estimations of the parameters are inferred. A typical compromise is to resort to the mode of the posterior distribution that is called the \emph{maximum a posteriori estimation} (MAP), described by
\begin{equation} \label{eq:map}
 \hat{\theta}_{\text{MAP}} = \arg\max_\theta p(\theta|w,X_s)
\end{equation}
Several calibration methods have been proposed and the large majority of them can be classified as probabilistic matching approaches. Some of the most commonly used calibration methods include weighted linear and non-linear least squares schemes, maximum likelihood estimates or Bayesian inference methods (e.g.  Markov Chain Monte Carlo, Particle and Kalman filters) \cite{AriasChao2015}. These methods differ in the level of complexity and the computational cost.

In this work, we propose an Unscented Kalman Filter \cite{Julier1997} to infer the values of the model correcting parameters $\hat{\theta}$ since our models of interest are nonlinear. However, the task can be also performed with other approaches. Hence, rather than focusing on one particular model calibration method, we evaluate the impact of different levels of calibration accuracy in the performance of the proposed fault detection and isolation algorithm and therefore in the framework proposed.

Model-based estimation of the sub-model health parameters from a transient data stream can be addresses with a traditional state-space formulation. In particular we consider an UKF where the state vector comprises the health parameters. The measurement equation depends on the states and the input signals at the present time step $t$; which is readily available from the system model $S$. Hence, we apply a UKF to a nonlinear discrete time system of the form:
\begin{align}
      \theta^{(t)} &= \theta^{(t-1)} + \xi^{(t)}  \\
       x_s^{(t)} &= S(w^{(t)}, \theta^{(t)}) + \epsilon^{(t)}
\end{align}
where $\xi \sim N(0,Q)$ is a Gaussian noise with covariance $Q$ and $\epsilon \sim N(0,R)$ is a Gaussian noise with covariance $R$. A more detailed explanation of this problem formulation applied to the monitoring of gas turbine engines can be found in Borguet \cite{Borguet2012}.  

The Kalman Filter provides estimates of $\hat{\theta}^{(t)}$ and therefore a fault detection and isolation (FDI) logic is also required. The standard approach for the FDI logic is to define thresholds for all $\hat{\theta}^{(t)}$ and rise an alarm if the inferred unobserved model parameter surpasses the defined threshold. However, such a approach has some limitations since a fault signature can be manifested only as a subtle change of the degradation rate. Accordingly, a threshold based approach will not be able to detect the presence of the fault signature until the fault is clearly manifested which will result in a detection delay. Similarly, a fault mode can result in different degradation rates in several of the monitored internal model parameters. As a result, fault signatures that are ambiguous when considered in each of the individual dimension of $\theta$ can be more clearly detected when considered combined. Therefore, we propose an alternative fault detection and isolation logic that is able to overcome some of the limitations of the threshold model-base approaches or a pure data-driven diagnostics models.

It should be noted that the proposed formulation of the calibration problem assumes that the system model has a good representation of the real physical process. This is a common situation when evaluating the health state of mature products where the system model has been developed and validated based on multiple field units or test beds units. In contrast, this is not the case of new developments. In the general, no model is perfect and a certain level of missing physical representation on the system model will imply in a lower calibration quality. In case of significant missing physics within the system model representation, the impact of model degradation gets entangled with the model correction rooted in the lack of physics. To mitigate this situation the calibration problem needs to be reformulated to account for a model discrepancy term $\delta(w)$ as follows:

\begin{align}
       X_s &= S(w, \theta(w)) + \delta(w) + \epsilon\\
       \theta(w) &= f(w) + \xi
\end{align}

Hence, the solution of this reformulated calibration problem involves finding the functions $\delta(w)$ and $\theta(w)$. The current state of the art solution is a sequential process. The equation is first solved for $\theta(w)$ and subsequently for $\delta(w)$. However, this approach is not optimal since smearing between the correction in $\theta$ and $\delta(w)$ is typically present in the solution. On the other hand, the simultaneous solution of $\theta(w)$ and $\delta(w)$ is an open research question that we do not address in this paper. 

\subsection{Fault Detection} \label{sec:detection-methods}
Several approaches for fault detection problems have been proposed in the literature. One of the main distinction criteria between them is the availability of labeled data. If labeled data from faulty and healthy operation are available, the problem is typically defined as a binary classification. However, faulty system conditions in critical systems are rare resulting in relatively few or even no faulty condition monitoring data. The focus of this paper is on the latter scenario, for which we define the problem as a \textit{one-class classification} \cite{Moya1996}.

\textbf{One-Class Classification}. The fault detection problem has been successfully addressed as a \textit{one-class classification} problem in \cite{Michau}. In this case the task turns to a regression problem that aims to discover a functional map $\mathbf{\mathcal{G}}$ from the healthy operation conditions to a target label $\mathbf{T} = \{h^{(i)} \; | \;  x^{(i)} \in S_{T}\}$ where $S_{T} \subsetneq \mathcal{D}_{L}$ is a training subset of $\mathcal{D}_{L}$. We consider a neural network model to discover the functional map $\mathbf{\mathcal{G}}$ and hence we refer to such a network as the \textit{one-class network}. The output of the one-class network will deviate from the target value $\mathbf{T}$ when the inner relationship of a new data point $x^{(j)} \in \mathcal{D}_{T+}$ does not correspond to the one observed in $S_{T}$. Therefore, we consider an unbounded similarity score $s_I(x^{(j)};\beta)$ of $x^{(j)}$ with respect to our healthy labeled data based on the absolute error of the prediction $\mathbf{\mathcal{G}}(x^{(j)})$ that we define as follows:

\begin{equation} \label{eq:similarity}
  s_I(x^{(j)};\beta) =  \frac{\mid \mathbf{T}- \mathbf{\mathcal{G}}(x^{(j)}) \mid}{\beta}
\end{equation}
\begin{equation}
  \beta =  P_{99.9}({\mid (\mathbf{T} - \mathbf{\mathcal{G}}(S_{V}) \mid})\gamma
\end{equation}

where $\beta$ corresponds to a normalizing threshold given by the 99.9\% percentile of the absolute error of the prediction of $\mathbf{\mathcal{G}}$ in a validation set (i.e $S_{V}$) extracted from $\mathcal{D}_{L}$ multiplied by a safety margin $\gamma=1.5$. Please note, that the percentile and $\gamma$ are hyper-parameters and can be adjusted to the specific problem. 

Hence, our fault detection algorithm is simply given by: 

\begin{equation} \label{eq:detection}
	\hat{h}_{s}(x^{(j)}) = 
	\begin{cases} 
		1 & s_I(x^{(j)}; \beta) < 1  \\
		0 & \text{otherwise} 
	\end{cases}
\end{equation}

To obtain the mapping function ${\mathcal{G}}$ we resort to a partially supervised learning strategy with embedding given only one target label $h_s^{(i)}=1$ at training.

\textbf{Partially Supervised One-Class Learning with Embedding}. The goal of a supervised learning strategy to discover a direct mapping from input $X$ to a target label $T$ given a training set $S_T$. An alternative strategy to this direct mapping is to obtain a representation of the raw input data (a.k.a. non-linear embedding) from which a reliable optimal mapping $\mathcal{G}$ can be learned. Hence, the task has two parts. Firstly, we find a transformation $E: {X}_L \longmapsto \mathbf{z_L}$ of the input signals to a latent space $\mathbf{z_L} \in R^{\,u \times d}$ that encode optimal distinctive features of ${X_L}$ in an unsupervised way (i.e. without having information on the labels). In a second step, we find a functional mapping $\mathcal{G}_{sle}: \mathbf{z}_{L} \longmapsto \mathbf{T}$ from the latent representation of input signals $\mathbf{z}_{L}=\{E(x^{(i)}) \; | \; x^{(i)} \in S_{T}\}$ to the label class $\mathbf{T}$. Since in our one-class problem formulation the training target contains only one class and since the number and nature of the fault classes in $\mathcal{D}_{T+}$ are not known in advance, we denote the corresponding supervised problem as partially supervised learning. This is a key difference to conventional supervised learning diagnostics where the available labeled (training) data samples already cover the essential information on the number and type of classes and the new observation only fall in the category of already known classes.

Different unsupervised deep-learning models can be considered to discover the latent representation $z_L$. In order to cover the most prominent deep neural network architectures and to show the performance independence of our proposed hybrid method to the network architectures we implemented two discriminative and one generative autoencoder variants. For the discriminative autonencodes, we considered vanilla autonecoders (AE) and hierarchical extreme learning machines (HELM) \cite{Zhu2015}. For the generative methods we implemented variational autoencoders (VAE) \cite{Kingma2014}. For the one-class network we use a discriminative model based on a feed-forward network (FF). A formal introduction to the selected neural networks model is provided in Section \ref{sec:AppendixII}.

It should be noted that our proposal for an embedding representation is not related to the quality of the one-class network to discriminate healthy and faulty conditions but to the need of performing fault isolation. The detection problem can also be formulated without an embedding (i.e. direct mapping from input $X$ to a target label $T$ given a training set $S_T$).

\subsection{Fault isolation} \label{sec:isolation-methods}
The autoencoder formulation of the problem allows to compute the expected signal values under the training distribution (i.e. $X$). The output of the autoencoder network $F(x^{(j)})$ will deviate from the input value $X$ when the inner relationship of a new data point $x^{(j)} \in \{\mathcal{D}_{U},\mathcal{D_{T}}\}$ does not correspond to the one observed in the training set $S_{T}$. Therefore, we compute the absolute deviation that each component of the reconstructed signals has (i.e. $|x_k^{(j)}- F(x^{(j)})_k|$) relative to the error observed in the validation dataset $S_V$ (i.e. healthy operation conditions).
\begin{equation} \label{eq:isolation}
  d_I(x_k^{(j)};\nu_k) =  \frac{|x_k^{(j)}- F(x^{(j)})_k|}{\nu_k}
\end{equation}
where $\nu$ corresponds to a normalizing threshold given by the 99.9\% percentile of the absolute error of the prediction of $F$ in the validation set $S_{V}$
\begin{equation}
  \nu_k = P_{99.9}\big(\{|x_k^{(j)} - F(x^{(i)})_k| \; | \; x_k^{(i)}\in S_V\}\big) 
\end{equation}
$d_I(x_k^{(j)};\nu)$ is an unbounded measure of similarity between the signal value predicted by the autoencoder network and the expected or true signal value. In our hybrid approach, the input space to the autoencoder comprises the calibration factors $\theta$ and the observed signals $X_s$ and therefore deviations in the signal reconstruction can be pointed out for measurement and model tuning factors.

\section{Case Study} \label{sec:Case_study}

\subsection{A Single Fault Mode in a Turbofan Engine} 
\begin{figure}[ht]
\centering
\includegraphics[width=8.5cm]{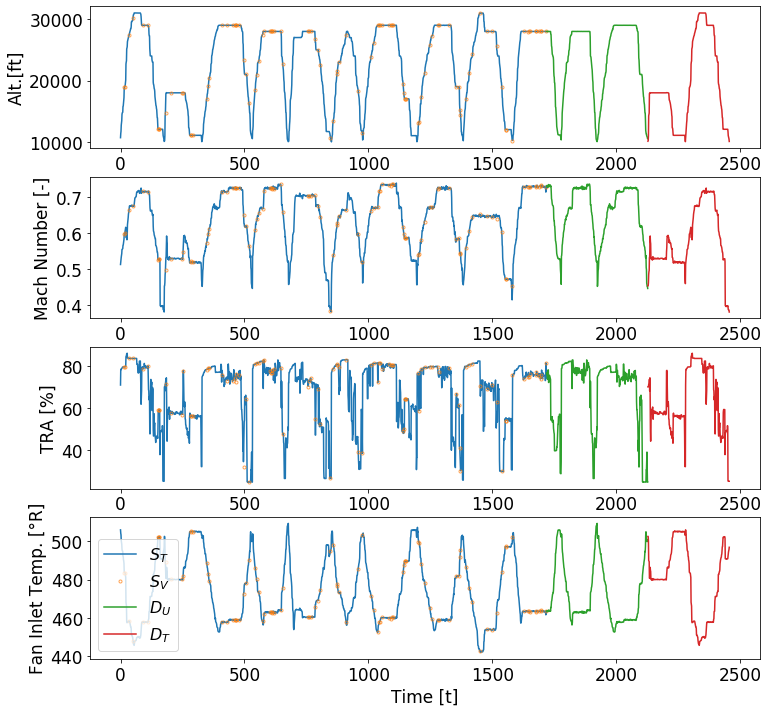}
\caption{Subset of 10 flight envelopes given by the traces of altitude (top), flight Mach number (middle) and throttle-resolver angle -TRA (bottom). Four dataset are shown: $S_T$ (blue), $S_V$ (orange), $D_U$ (green) and $D_T$ (red).}
\label{fig:operation_profile}
\end{figure}

A new dataset was designed to evaluate the proposed methodology. The CMAPSS dataset $\text{DS00}$ provides simulated condition monitoring data of an advanced gas turbine during 24 flights cycles. The dataset was synthetically generated with the Commercial Modular Aero-Propulsion System Simulation (C-MAPSS) dynamical model \cite{Frederick2007}. Real flight conditions as recorded on board of a commercial jet were taken as input to the C-MAPSS model \cite{DASHlink}. Figure \ref{fig:operation_profile} shows 14 simulated flight envelopes given by the traces of altitude ($\text{alt}$), flight Mach number ($\text{XM}$), throttle-resolver angle ($\text{TRA}$) and total temperate at the fan inlet ($\text{T2}$). Each flight cycle contains $\sim$175 snapshots of recordings covering climb, cruise and descend flight conditions (i.e. $\text{alt}>10000$ ft). The labeled dataset $\mathcal{D}_L$ (blue) consists of 20 flight cycles with a healthy state of the engine (i.e $h_s=1$). The unlabeled and test datasets $\{\mathcal{D}_U, \mathcal{D_{T}}\}$ (green and red respectively) contain snapshots of $R=4$ concatenated flight cycles with a deteriorating engine condition. The intensity of the degradation increases at each flight (i.e  step-wise increase). The fault mode corresponds to a high pressure compressor (HPC) efficiency degradation. Each of the fault magnitudes is denoted with a fault id (see Table \ref{tb:failures}). The unlabeled dataset also includes 60 snapshots of initial healthy operation. The unlabeled and test datasets $\{\mathcal{D}_U, \mathcal{D_{T}}\}$ contain a subset of flight conditions experienced during training. Therefore, this set-up relates to a real scenario where an aircraft is operating under certain flight routes, which results in very similar flight condition. In addition to the noisy flight conditions, all the healthy operative conditions incorporate white noise in all the engine health model parameters (see Table \ref{tb:theta}). No additional noise or bias is considered for sensor readings. A total of $\sim 3200$ healthy data points are available for training. The unlabeled and test datasets $\{\mathcal{D}_U, \mathcal{D_{T}}\}$ contain $\sim 740$ data points.

\begin{table}[ht]
\begin{center}
\begin{tabular}{clccc} \hline
Fault Id         & Fault Mode       &  Magnitude   & dataset         \\\hline
1                & HPC Efficiency  &  -0.5 \%      & $\mathcal{D}_U$   \\
2                & HPC Efficiency  &  -1.0 \%      & $\mathcal{D}_U$   \\ 
3                & HPC Efficiency  &  -1.5 \%      & $\mathcal{D}_T$   \\
4                & HPC Efficiency  &  -2.0 \%      & $\mathcal{D}_T$   \\ \hline
\end{tabular}
\end{center}
\caption[Table tb:failures]{Overview of the generated faults}
\label{tb:failures}
\end{table}

\subsection{Pre-processing} \label{sec:pre-processig}
The dimension of the input space $X$ (i.e. n) varies depending on the solution strategy considered (see Table \ref{tb:input}). The diagnostic model based on condition monitoring data (CMBD) has 17 inputs. The residual-based model considers 31 inputs and the proposed hybrid method based on system model calibration (CBHD) uses to 45 inputs. Tables \ref{tb:W_X_m} to \ref{tb:delta} provide a detailed overview of the model variables included in the condition monitoring signals $[W, X_s]$, model residuals $\delta_{X_s}$, virtual sensors $\hat{X}_{v}$ and model calibration parameters $\hat{\theta}$.

\begin{table}[ht]
\begin{center}
\begin{tabular}{llc} \hline
      Model             &   Input                                           &  n     \\  \hline
      CMBD - No Hybrid  &  $[W, X_s]$                                       &  17    \\
      Residual          &  $[W, \delta_{X_s}]$                              &  31    \\
      CBHD - Hybrid     &  $[W, \hat{X}_{s}, \hat{X}_{v}, \hat{\theta}]$    &  45    \\  \hline
\end{tabular}
\end{center}
\caption[Table caption text]{Dimension of the input space for the autoencoder network - $n$}
\label{tb:input}
\vspace{-0mm}
\end{table}

\begin{table}[ht]
\begin{center}
\begin{tabular}{cllc}
\hline
Id            & Symbol       &  Description                       & Units          \\ \hline
1             & alt          & Altitude                           & ft             \\
2             & XM         & Flight Mach number                 & -              \\
3             & TRA          & Throttle-resolver angle            & \%             \\
4             & Wf           & Fuel flow                          & pps            \\
5             & Nf           & Physical fan speed                 & rpm            \\
6             & Nc           & Physical core speed                & rpm            \\
7             & T2           & Total temperature at fan inlet     & $^{\circ}$R    \\
8             & T24          & Total temperature at LPC outlet    & $^{\circ}$R    \\
9             & T30          & Total temperature at HPC outlet    & $^{\circ}$R    \\
10            & T48          & Total temperature at HPT outlet    & $^{\circ}$R    \\
11            & T50          & Total temperature at LPT outlet    & $^{\circ}$R    \\
12            & P15          & Total pressure in bypass-duct      & psia           \\
13            & P21          & Total pressure at fan outlet       & psia           \\
14            & P24          & Total pressure at LPC outlet       & psia           \\
15            & Ps30         & Static pressure at HPC outlet      & psia           \\
16            & P40          & Total pressure at burner outlet    & psia           \\
17            & P50          & Total pressure at LPT outlet       & psia           \\ \hline
\end{tabular}
\end{center}
\caption[Table tb:CM]{Condition monitoring signals - $[W,X_s]$. The Id is used in this document as shorthand of the variable description. The variable symbol corresponds to the internal variable name in CMAPSS. The descriptions and units are reported as in the model documentation \cite{Frederick2007}.}
\label{tb:W_X_m}
\end{table}

\begin{table}[ht]
\begin{center}
\begin{tabular}{cllc}
\hline
Id  & Symbol       &  Description                       & Units         \\ \hline
18  & T40          & Total temp. at burner outlet       & $^{\circ}$R   \\
19  & P30          & Total pressure at HPC outlet       & psia    \\
20  & P45          & Total pressure at HPT outlet       & psia    \\
21  & W21          & Fan flow                           & pps     \\
22  & W22          & Flow out of LPC                    & lbm/s   \\
23  & W25          & Flow into HPC                      & lbm/s   \\
24  & W31          & HPT coolant bleed                  & lbm/s   \\
25  & W32          & HPT coolant bleed                  & lbm/s   \\
26  & W48          & Flow out of HPT                    & lbm/s   \\
27  & W50          & Flow out of LPT                    & lbm/s   \\
28  & epr          & Engine pressure ratio (P50/P2)     & --      \\
29  & SmFan        & Fan stall margin                   & --      \\
30  & SmLPC        & LPC stall margin                   & --      \\
31  & SmHPC        & HPC stall margin                   & --      \\
32  & NRf          & Corrected fan speed                & rpm     \\
33  & NRc          & Corrected core speed               & rpm     \\
34  & PCNfR        & Percent corrected fan speed        & pct     \\
35  & phi          & Ratio of fuel flow to Ps30         & pps/psi \\ \hline
\end{tabular}
\end{center}
\caption[Table tb:failures]{Virtual sensors - $[X_v]$. The Id is used in this document as shorthand of the variable description. The variable symbol corresponds to the internal variable name in CMAPSS. The descriptions and units are reported as in the model documentation \cite{Frederick2007}.}
\label{tb:X_v}
\end{table}

\begin{table}[ht]
\begin{center}
\begin{tabular}{cllc} \hline
Id & Symbol         &  Description                       & Units    \\\hline
36  & fan\_eff\_mod  & Fan efficiency modifier            & -       \\
37  & fan\_flow\_mod & Fan flow modifier                  & -       \\
38  & LPC\_eff\_mod  & LPC efficiency modifier            & -       \\
39  & LPC\_flow\_mod & LPC flow modifier                  & -       \\
40  & HPC\_eff\_mod  & HPC efficiency modifier            & -       \\
41  & HPC\_flow\_mod & HPC flow modifier                  & -       \\
42  & HPT\_eff\_mod  & HPT efficiency modifier            & -       \\
43  & HPT\_flow\_mod & HPT flow modifier                  & -       \\
44  & LPT\_eff\_mod  & LPT efficiency modifier            & -       \\
45  & LPT\_flow\_mod & HPT flow modifier                  & -       \\ \hline
\end{tabular}
\end{center}
\caption[Table tb:failures]{Model correcting parameters - $[\theta]$. The Id is used in this document as shorthand of the variable description. The variable symbol corresponds to the internal variable name in CMAPSS. The descriptions and units are reported as in the model documentation \cite{Frederick2007}.}
\label{tb:theta}
\end{table}

\begin{table}[ht]
\begin{center}
\begin{tabular}{cllc}
\hline
Id            & Symbol                &  Description                             & Units       \\ \hline
$\delta_4$    & $\delta_{Wf}$         & Delta fuel flow                          & pps         \\
$\delta_5$    & $\delta_{Nf}$         & Delta physical fan speed                 & rpm         \\
$\delta_6$    & $\delta_{Nc}$         & Delta physical core speed                & rpm         \\
$\delta_7$    & $\delta_{T2}$         & Delta total temp. at fan inlet           & $^{\circ}$R \\
$\delta_8$    & $\delta_{T24}$        & Delta total temp. at LPC outlet          & $^{\circ}$R \\
$\delta_9$    & $\delta_{T30}$        & Delta total temp. at HPC outlet          & $^{\circ}$R \\
$\delta_{10}$ & $\delta_{T48}$        & Delta total temp. at HPT outlet          & $^{\circ}$R \\
$\delta_{11}$ & $\delta_{T50}$        & Delta total temp. at LPT outlet          & $^{\circ}$R \\
$\delta_{12}$ & $\delta_{P15}$        & Delta total press. in bypass-duct        & psia    \\
$\delta_{13}$ & $\delta_{P20}$        & Delta total press. at fan inlet          & psia    \\
$\delta_{14}$ & $\delta_{P24}$        & Delta total press. at LPC outlet         & psia    \\
$\delta_{15}$ & $\delta_{Ps30}$       & Delta static press. at HPC outlet        & psia    \\
$\delta_{16}$ & $\delta_{P40}$        & Delta total press. at burner outlet      & psia    \\
$\delta_{17}$ & $\delta_{P50}$        & Delta total press. at LPT outlet         & psia    \\ \hline
\end{tabular}
\end{center}
\caption[Table tb:delta]{Delta to healthy state - $[\delta_{X_s}]$. The Id is used in this document as shorthand of the variable description. The variable symbol corresponds to the internal variable name in CMAPSS. The descriptions and units are reported as in the model documentation \cite{Frederick2007}.}
\label{tb:delta}
\end{table}

The input space $X$ to the models is normalized to a range $[-1, 1]$ by a min/max-normalization.  A validation set $S_{T} \subsetneq D_L$ comprising 6 \% of the labelled healthy data for all the models was chosen.

\subsection{Network architectures} \label{sec:networks}
\begin{figure}[ht]
\centering
\includegraphics[width=8.8cm]{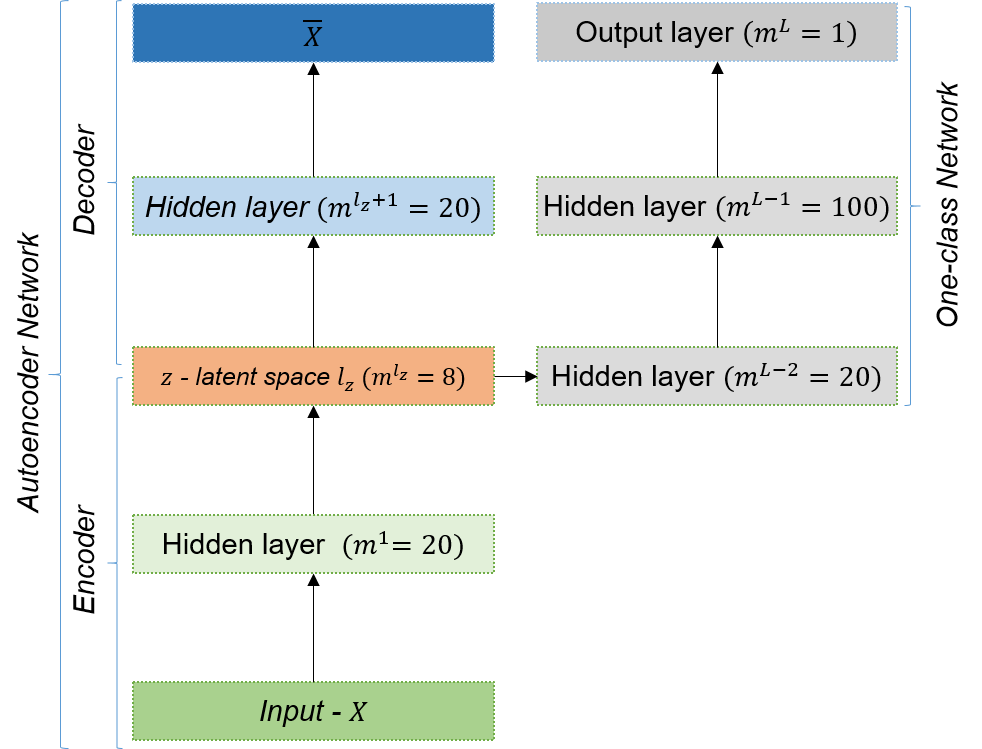}
\caption{Network architecture for the defined learning problem with an autoencoder (encoder-decoder) and the one-class detection network.}
\label{fig:network}
\end{figure}

The partially supervised with embedding learning strategies require an \textit{autoencoder network} in addition to the \textit{one-class network}. As shown in Figure \ref{fig:network}, the input signals $X$ are reconstructed by the encoder-decoder networks. The encoder provides a new representation $z$ of the input signals. The mapping to the target label $\mathbf{T}$ is carried out by the \emph{one-class network} taking as input the latent (i.e. unobserved) representation of the input data $z$.

To evaluate the different methods in a fair way, we separate the effect of regularization in the form of model and learning strategies choice from other inductive bias in the form of choice of neural network architecture. Therefore, we define a common architecture of the \textit{one-class network} and \textit{autoencoder network} for all our deep autoencoders.

\vspace{10mm}

\textbf{One-Class Network}. The proposed network topology uses three fully connected layers ($L=3$). The first hidden layer has 20 neurons (i.e. $m^{1}=20$), the last hidden layer has 100 neurons ($m^{L-1}=100$). The network ends with a linear output neuron ($m^{L}=1$). Therefore, in compact notation, we refer to the \textit{one-class network} architecture as $[20, 100, 1]$. \textit{tanh} activation function is used throughout the network. It should be noted that the one-class classification problem formulation is a regression problem and therefore the last activation $\sigma^L=I$ is the identity. 

\textbf{Autoencoder Networks}. Based on the same argument as mentioned above, the autoencoder models (i.e. AE and VAE) use the same encoder architecture with two hidden layers ($l_z=2$) with $m^1=20$ and latent space of 8 neurons ($d=8$). In compact notation, we refer to the \textit{autoencoder network} architecture as $[n, 20, 8, 20, n]$. Where $n$ denotes the size if the input space $X$; which varies depending on the solution strategy considered. The VAE model uses the mean of the approximate posterior (i.e. $\mu$) as the model latent space to avoid using approximate samples from posterior distribution (i.e. $z^{(i)}$). The HELM model reproduces the encoder and the one-class networks in one single hierarchical network. Hence, the resulting network architecture is $[n, 20, 8, 20, 100, 1]$.


\textbf{OC-SVM model}.To evaluate if a deep learning architecture is required for the fault detection task, we compare the results to the standard one-class support vector machines (One-Class SVM) \cite{Scholkopf2000} for novelty detection. This enables us to evaluate the benefits and the potential need for complex neural network architectures for the defined fault detection task. We use the standard $\textit{scikit-learn}$ \cite{scikit-learn} implementation of the one class SVM with an radial basis function kernel, $\textit{nu=0.001}$ and $\textit{gamma=0.1}$. The model performance is sensitive to the choice of these hyperparameters. Therefore, optimal parameters for the validation set $S_V$ may not guarantee a good performance in $D_T$. We selected the hyperparameters that maximize the $F_1$ score on the test set to ensure the best possible performance of the baseline on the test dataset. For other algorithms, the parameters were selected based on the validation dataset. This makes the comparison of the deep learning algorithms to the baseline even more challenging. 

\subsection{Training Set-up}
The optimization of the networks' weights of all the models was carried out with mini-batch stochastic gradient descent (SGD) and with the \textit{Adam} algorithm \cite{Kingma2014Adam}. \textit{Xavier} initializer \cite{Glorot} was used for the weight initializations. The learning rate (LR), epoch and batch size were set according to Table \ref{tb:Settings}. The batch size for the autonecoder network was set to 512 and to 16 for one-class network. Similarly, the number of epochs for autoencoder training was set to 2000 and for the supervised models to 500.  Therefore, all these methods use the same network architecture and hyper-parameters for the optimisation.

\begin{table}[ht]
\begin{center}
\begin{tabular}{lcccc} \hline
       Model         & LR       & Batch Size &  Epochs         \\ \hline
       One-class     & 0.001    &  16       &    500           \\
       Autoencoder   & 0.001    &  512      &    2000          \\ \hline
\end{tabular}
\end{center}
\caption[Table caption text]{Training parameters}
\label{tb:Settings}
\vspace{-0mm}
\end{table}

\subsection{Evaluation Metrics}
In order to compare and analyse the performance of our models on the intended diagnostics task we defined two evaluation aspects: detection of unknown faults (i.e. estimation of $h_s$) and fault isolation. For each of the two aspects, we consider targeted evaluation metrics that are defined in the following.

\textbf{Fault Detection}. Given the combined test dataset $D_{T+}$ with true health state $h_s^{(j)}$ and the corresponding estimated health state $\hat{h}_s^{(j)}$, we evaluate the performance of the fault detection algorithm as the accuracy of a binary classification problem
\begin{align} \label{eq:classmetric}
\text{Acc} &= \frac{1}{M+m}\sum_{i=1}^{M+m} \mathbf{1}(h_s^{(j)} = \hat{h}_s^{(j)})
\end{align}
where $M+m$ number of data points in $D_{T_{+}}$ and $\mathbf{1}\{.\}$ denotes the indicator function.

\textbf{Fault Isolation}. The error in the reconstruction signal will be more notorious for those signals in close relation to the fault root cause. Therefore, we report the index of the signals of those components of the data point $x^{(j)}$ that satisfy $d_I(x_k^{(j)};\nu_k)>1$. The mapping between variable description of the variables and the corresponding index is provided in Tables \ref{tb:W_X_m} to \ref{tb:delta}.

\section{Experimental Results} \label{sec:Results}

\subsection{Fault Detection} \label{sec:Results-fd}

Table \ref{tb:detection} shows the performance of our twelve models on fault detection. The residual-based and the calibration-based approaches achieve nearly $100\%$ detection accuracy independently of the neural network model considered. Both approaches provide an improvement of nearly $80\%$ with respect to the best diagnostic model based purely on condition monitoring data (i.e. \emph{AE} with $[W, X_s]$). The OC-SVM model results in a lower detection accuracy than the deep learning models with independence on the input space considered. Concretely, the best performing autoencoder model (i.e. AE) provides c.a. 4\% accuracy improvement with respect to the standard OC-SVM when a calibration-based approaches is considered.

\begin{table}[ht]
\begin{center}
\begin{tabular}{lcccc}  \hline
      Input                                         &  AE              & VAE      & HELM    &  OC-SVM   \\  \hline
      $[W, X_s]$                                    &  24.5            & 12.9     & 8.0     &  10.0     \\
      $[W, \delta_{X_s}]$                           &  98.7            & 99.3     & 8.5     &  79.5     \\
      $[W, \hat{X}_{s}, \hat{X}_{v}, \hat{\theta}]$ & $\mathbf{100.0}$ & 99.1     & 97.5    &  96.2     \\  \hline      
\end{tabular}
\end{center}
\caption[Table caption text]{Overview of detection results. Mean values of 10 runs}
\label{tb:detection}
\end{table}

\begin{figure}[ht]
\centering
\includegraphics[width=8.5cm]{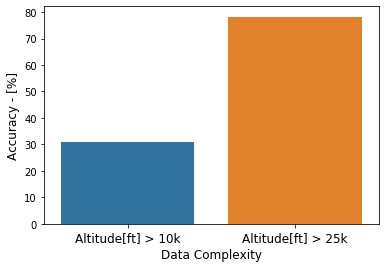}
\caption{Evolution of the accuracy with dataset complexity for AE model based on $[W, X_s]$ inputs in faults 1, 2 and 4. Fault 3 is not considered since $alt < 25000$ ft.}
\label{fig:complexity}
\end{figure}

Diagnostics models based on condition monitoring data show poor performance independently on the autoencoder network considered. A possible explanation for this may be that this is due to the high complexity of the dataset in the form of a large variability in the input space due to varying operating conditions. To verify this idea we trained the \textit{AE} model with CM inputs on a subset of the training data with operative points closer to cruise conditions. Hence, we restricted the fight altitude to above $25000$ ft. Figure \ref{fig:complexity} shows the accuracy of the diagnostics model based on condition monitoring data trained in this simpler dataset. We can observe that the detection performance drastically increases; which supports our hypothesis.

\begin{figure}[ht]
\centering
\includegraphics[width=8cm]{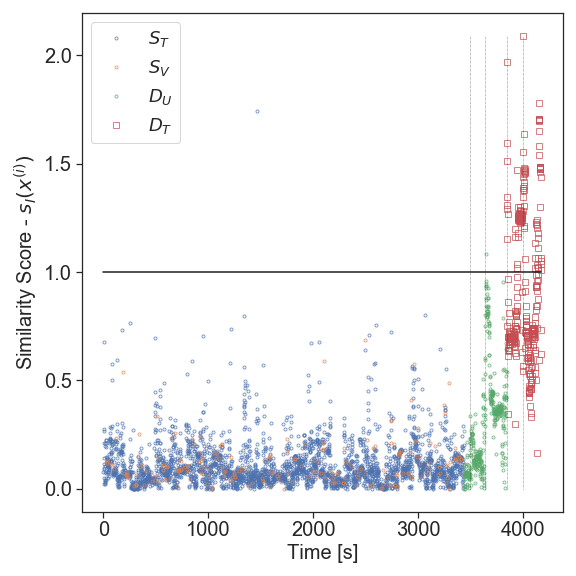}
\caption{Similarity index for four HPC efficiency faults of different intensities with AE model based on $[W, X_s]$ signals. All the faults occur at different flight conditions. The decision threshold is plotted as horizontal black line ($s=1$). The onset times of each fault are indicated by the vertical dashed lines. Four dataset are shown: $S_T$ (blue), $S_V$ (orange), $D_U$ (green) and $D_T$ (red).}
 \label{fig:s_score}
\end{figure}

The detection performance of the one-class solutions reported in Table \ref{tb:detection} is determined by the capability of the similarity score $s_I(x^{(i)};\beta)$ to represent a valid and consistent distance to healthy operation learnt in the training phase (i.e. $D_L$). To demonstrate and verify this behaviour we plot in Figure \ref{fig:s_score} the similarity score obtained with \textit{AE} model with CM inputs in the four HPC efficiency faults of increasing intensities (-0.5\% to -2\%). The onset of each fault is indicated by the dashed vertical lines. We observe that the more severe the fault is, the higher the detection index. Therefore, $s(x^{(i)};\beta)$ shows the expected consistency. However, we also observe that only for HPC faults with intensities below -1.0\% the similarity score is above the decision threshold $s(x^{(i)};\beta)>1$ (black horizontal line). Hence, the one-class network fails to discriminate between healthy and faulty conditions for HPC efficiency deterioration below 1.0\%. 

\begin{figure}[ht]
\centering
\includegraphics[width=8.5cm]{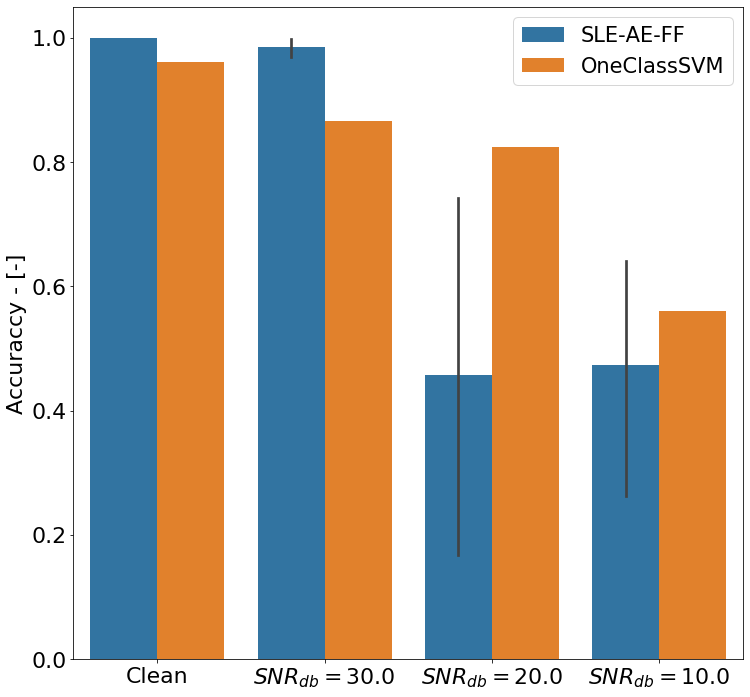}
\caption{Fault detection accuracy as function of the noise levels for AE and OC-SVM model. $95\%$ confidence intervals are shown as rectangular bars.}
 \label{fig:noise_impact}
\end{figure}

\begin{figure}[ht]
\centering
\includegraphics[width=8.5cm]{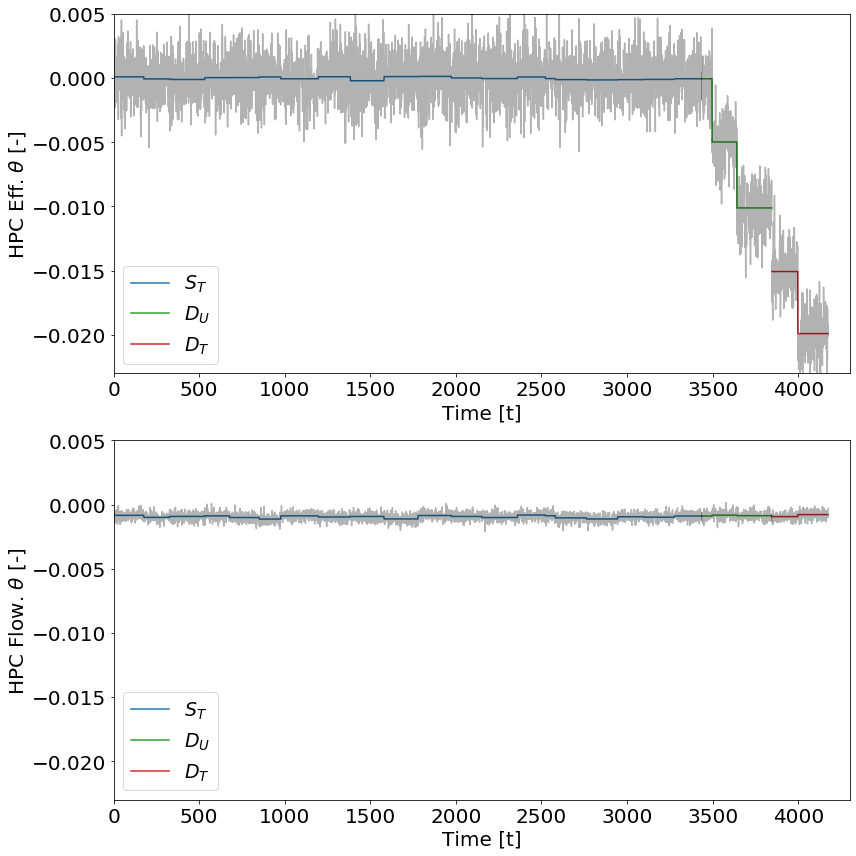}
\caption{Noisy calibration factors for a noise level of $\text{SNR}_{db}=10$ imposed on the high pressure compressor (HPC) efficiency. The added noise to the high pressure compressor (HPC) flow (which is not affected by the fault mode) is shown as reference. Three datasets are shown: $S_T$ (blue), $D_U$ (green) and $D_T$ (red). The cycle of transition from \emph{normal degradation} to \emph{abnormal degradation} is indicated by the dotted vertical line}
 \label{fig:noise_signal}
\end{figure}

The quality of the calibration process has an impact on the fault detection performance. In order to quantify this impact, the calibration factors are contaminated with noise of different signal-to-noise ratios. We impose the noise perturbation to all the calibration factors (i.e $\theta \in R^{10}$), however the impact is more pronounced for \emph{HPC\_Eff\_mod} since it defines the fault mode. Figure \ref{fig:noise_signal} shows two components of the resulting noisy calibration process. Figure \ref{fig:noise_impact} shows the impact in fault detection accuracy of the different noise levels for two best performing models. We can observe a decrease in the accuracy as the noise increases for all the tested models. \text{OC-SVM} model shows the most robust performance for noise levels $\text{SNR}_{db}<30$. Most of the models are able to achieve an accuracy that is above the pure data-driven models if all the SNR are evaluated. Therefore, these results demonstrate the robustness of the proposed fault detection approach. It should be noted that the $\text{SNR}_{db}$ scale is logarithmic.

\subsection{Fault Isolation} \label{sec:Results-fi}

Table \ref{tb:isolation} shows the input signals detected as anomalous with the AE and VAE models. For simplicity, we report the index of the signals according to Tables \ref{tb:W_X_m} to \ref{tb:delta}. The affected signals are presented in a decreasing order according the value of the similarity indicator $d_I(x_k^{(j)};\nu_k)$. Hence, the most affected signals are presented first. Only variables that satisfy $d_I(x_k^{(j)};\nu_k)>1$ are reported.

The four faults present in the combined test sets $\mathcal{D}_{T+}$ are rooted in a HPC efficiency deficit. However, not all the models have an input space where the compressor efficiency is represented. Concretely, only the calibration-based hybrid model with inputs $[W, \hat{X}_{s}, \hat{X}_{v}, \hat{\theta}]$ has a representation of the HPC efficiency through the estimated model correcting parameters $\hat{\theta}$. Therefore, in the best case, the remaining models can only aim to place the root cause of a HPC degradation on variables physically related to the HPC. For instance, models that consider only condition monitoring signals $[W, X_s]$ detect a large reconstruction error in variable 6 (i.e. the rotational core speed of the shaft where the high pressure compressor is placed). The hybrid model based on residual $[W, \delta_{X_s}]$ encodes the fault signature in five residuals: $\delta_{11}$, $\delta_{10}$, $\delta_{9}$, $\delta_{6}$ and $\delta_{8}$. Therefore, the residual of core speed $\delta_6$ is also detected as an affected signal in addition to the HPC outlet temperature ($\delta_9$) and temperatures at the outlets of the High and Low Pressure Turbines (i.e. $\delta_{10}$ and $\delta_{11}$). The isolation of these last two process variables as the fault root cause is a clear smearing of the effect of an HPC degradation to other unrelated subsystems. Neural networks based on VAE show a similar isolation performance.

Finally, hybrid models based on calibrated models with input signals $[W, \hat{X}_{s}, \hat{X}_{v}, \hat{\theta}]$ encode the fault signature in only variable 40; which corresponds to the component of $\theta$ representing the correction of the HPC efficiency. Any model with $[W, \hat{X}_{s}, \hat{X}_{v}, \hat{\theta}]$ provides perfect isolation.

%
\begin{table}[ht]
\begin{center}
\begin{tabular}{|l||c|c|c|c|}      \hline
      \multicolumn{5}{|c|}{AE}  \\ \hline \hline
      Input                                          & -0.5\%                               & -1.0\%       & -1.0\%     & -2.0\%                \\ \hline
      $[W, X_s]$                                     & \multicolumn{2}{|c|}{-}                             & 6          & 6, 15      \\ \hline
      $[W, \delta_{X_s}]$                            &  \multicolumn{4}{c|}{$\color{red}\delta_{11}$, $\color{red}\delta_{10}$, $\delta_{9}$, $\delta_{6}$, $\color{red}\delta_{8}$}   \\  \hline
      $[W, \hat{X}_{s}, \hat{X}_{v}, \hat{\theta}]$  &  \multicolumn{4}{c|}{40}                                                      \\ \hline
      \multicolumn{5}{|c|}{VAE}  \\   \hline \hline
      Input                                          & -0.5\%                               & -1.0\%       & -1.0\%     & -2.0\%     \\ \hline
      $[W, X_s]$                                     &  \multicolumn{4}{|c|}{-}                                                      \\ \hline
      $[W, \delta_{X_s}]$                            &  \multicolumn{4}{c|}{$\color{red}\delta_{10}$, $\color{red}\delta_{11}$, $\delta_{6}$, $\delta_{9}$, $\color{red}\delta_{8}$}  \\ \hline
      $[W, \hat{X}_{s}, \hat{X}_{v}, \hat{\theta}]$  &  \multicolumn{4}{c|}{40}                                                      \\ \hline
    \end{tabular}
  \end{center}
\caption[Table caption text]{Overview of isolation results on four HPC efficiency faults with impact from -0.5\% to -2.0\%. The table shows the index of the affected variables as introduced in Tables \ref{tb:X_v}-\ref{tb:delta}. Variables affected by smearing are colored in red.}
\label{tb:isolation}
\end{table}

\subsection{Feature Representation} \label{sec:Results-fr}

The results presented in the previous section have demonstrated that the proposed hybrid approach provides a very good performance for fault detection and isolation, particularly for systems with a high variability of the operating conditions. To better understand how the different (expanded) input spaces affect the latent representation and also the performance of the models on the diagnostics tasks, the latent space of the different models is analyzed. Please note that the analysis of the latent space is mainly performed for understanding and demonstration purposes. Therefore, only the first two dimensions of the latent space are visualized. While this does not provide a full evaluation of the latent space, a separability of the healthy and faulty conditions in the first two dimensions of the latent space would support the assumption that such a representation would also be favorable for the diagnostics tasks based on this latent representation. 

\begin{figure}[htb]
\centering
\includegraphics[width=8.6cm]{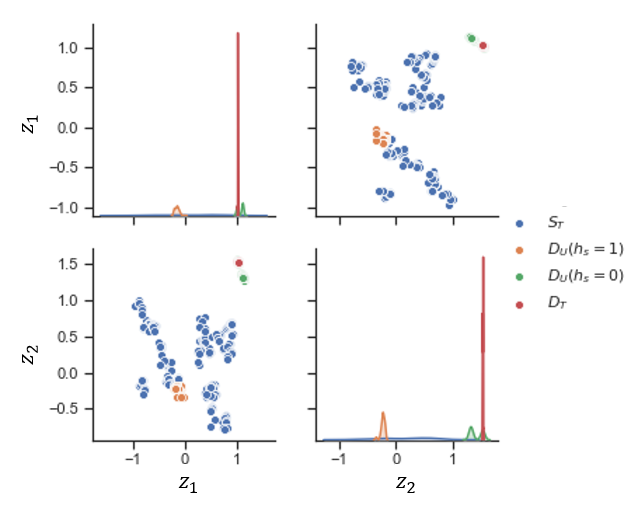}
\caption{Pairwise scatter plot the first two components of the latent space $z$ of the hybrid AE model with $X=[W, \hat{\theta}, \hat{X}_{s}, \hat{X}_{v}]$. The scatter plot is colored according to the dataset of origin: $S_T$ (blue), $D_U(h_s=1)$ (orange), $D_U(h_s=0)$ (green) and $D_T$ (red)}
\label{fig:features_x4}
\end{figure}

\begin{figure}[hbt]
\centering
\includegraphics[width=8.6cm]{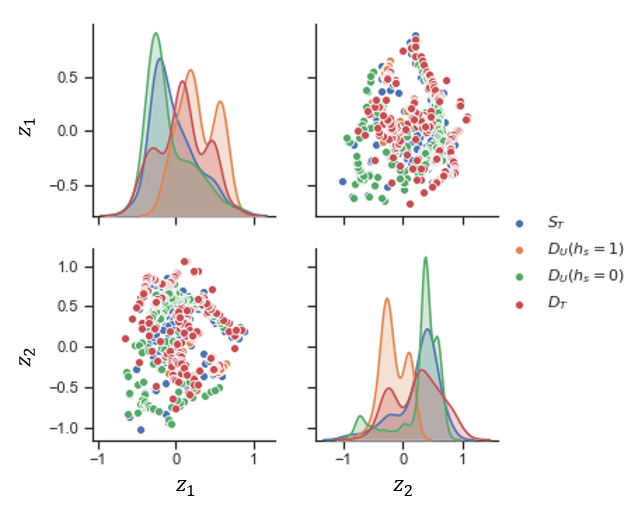}
\caption{Pairwise scatter plot the first two components of the latent space $z$ of the data-driven AE  with  $X= [W, X_s]$. The scatter plot is colored according to the dataset of origin: $S_T$ (blue), $D_U(h_s=1)$ (orange), $D_U(h_s=0)$ (green) and $D_T$ (red)}
\label{fig:features_x1}
\end{figure}

\begin{figure}[hbt]
\centering
\includegraphics[width=8.6cm]{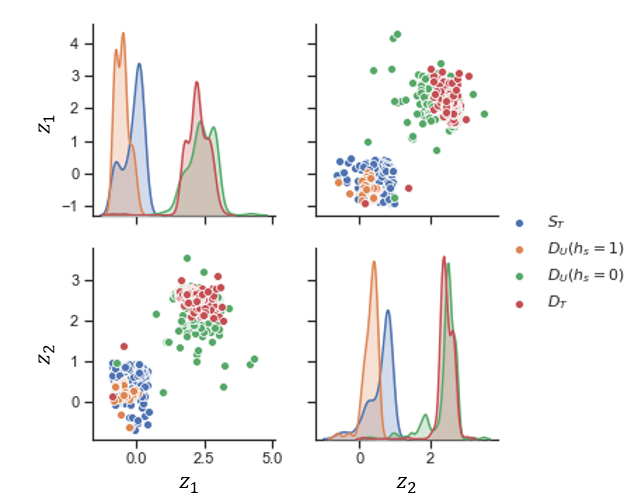}
\caption{Pairwise scatter plot the first two components of the latent space $z$ of the hybrid delta AE model with $X= [W, \delta_{X_s}]$. The scatter plot is colored according to the dataset of origin: $S_T$ (blue), $D_U(h_s=1)$ (orange), $D_U(h_s=0)$ (green) and $D_T$ (red)}
\label{fig:features_x2}
\end{figure}

Figure \ref{fig:features_x4} shows a pairwise scatter plot of the first two dimensions of the latent space $z$ of the hybrid AE model $X=[W, \hat{X}_{s}, \hat{X}_{v}, \hat{\theta}]$, while Figure \ref{fig:features_x1} represents the first two dimensions of the latent space of the data-driven model $X= [W, X_s]$. The latent space of the residual-based approach ($[W, \delta_{X_s}]$) is represented in Figure \ref{fig:features_x2}. The scatter plots are colored according to the dataset of the origin. The healthy class (i.e $S_T$) is shown in blue, healthy unlabelled data from $D_U(h_s=1)$ are shown in orange and the faulty operative conditions from the test set $D_T$ are represented in green.

It can be clearly observed that expanding the input space with additional model variables has a large impact in the latent representation. Concretely, the faulty conditions are clearly clustered together and have a high distance to the healthy operating conditions (centered around zero) for the two hybrid approaches. On the contrary, a distinction between healthy and faulty conditions in the latent representation of the purely data-driven approach $X= [W, X_s]$ is not possible. The representation of healthy and unhealthy classes shows clear overlaps in the two represented dimensions. These exemplary plots support the argument that the hybrid approaches provide a more favorable and more distinct representation of the healthy respectively unhealthy conditions. This results in a easier detection task of the one-class network leading to better detection results.


\section{Discussion} \label{sec:Discussion}
The performed experiments on the C-MAPSS dataset demonstrate that the proposed hybrid deep learning-based diagnostics algorithm, combining information from a physics-based model and condition monitoring data, outperforms pure data-driven deep learning methods in fault detection and isolation, particularly for systems with a high variability of operating conditions.

Standard residual-based approaches ($\delta_{X_{s}}$) and hybrid approaches based on calibrated model variables ($ \hat{X}_{s}, \hat{X}_{v}, \hat{\theta})$ led to similar fault detection performance. The analysis of the encoded representations showed that their excellent detection performance is rooted in the same concept. Both latent spaces provide a clear discrimination between healthy and faulty operating conditions; which simplifies the fault detection task. This result implies that an accurate model calibration is not relevant to obtain good detection performance as long as the system degradation or fault signature is encoded in model inferred variables (i.e. $\hat{\theta}$, $\delta X_{s}$ or both).

However, accurate fault isolation (overcoming the smearing effect) is only possible when model tuning parameters $\hat{\theta}$ are considered. Hence, the proposed hybrid approach based on calibrated inputs provides clear benefits for the fault isolation task.  However, it should be noted that this approach introduces an additional pre-processing step. Also, the performance of this approach depends on the calibration capabilities and it is expected that if the calibrated model fails to reproduce closely the reality, the capability to clearly isolate failures will decrease. 

Residual based and calibration based frameworks are not mutually exclusive and therefore a third option is to combine them. In this case, in addition to a pre-processing calibration step, the residuals $\delta X_s$ to a healthy system state are also computed. Hence, the input to the deep-learning diagnostics model would comprise $[w, \hat{X}_s, \hat{X}_v, \hat{\theta}, \delta_{X_s}]$. 





\section{Conclusions} \label{sec:Conclusion}

In this paper, we proposed a hybrid fault diagnosis framework combing the physical performance models with deep learning algorithms. 

The performance of the proposed framework was evaluated on a synthetic dataset generated with the Commercial Modular Aero-Propulsion System Simulation (C-MAPSS) dynamical model. The C-MAPSS dataset D00 provides simulated condition monitoring data of an advanced gas turbine during real flight conditions under healthy and four faulty operative condition. 

The proposed framework and method was able to outperform purely data-driven deep learning algorithms and the traditional OC-SVM model for fault detection (providing a perfect detection accuracy) and for fault isolation (being able to precisely isolate the root cause of the originating fault). The proposed methodology is able to overcome the smearing that is commonly observed in purely data-driven approaches where all the affected signals and not the root cause are isolated by the algorithms. 

More importantly, we showed that the advantages of hybrid models are particularly relevant for complex datasets with a large variability in the operating conditions. Under these conditions, purely data-driven deep-learning approaches derived from condition monitoring data fail to obtain a robust diagnostic model. However, for systems with more homogeneous operating conditions, we expect a similar performance between the hybrid and the data-driven approaches for fault detection tasks.  

A feature learning analysis indicates that the excellent detection results obtained with hybrid methods are rooted in the fact that the latent space $z$ provides a representation of the input signals that is clearly informative about the true label class. 

As demonstrated in the experiments, accurate isolation results are obtained when the calibrated system model has a good representation of the fault modes. However, the analysis of fault modes that are not represented in the system model is of interest for practical applications. In this situation, it could be expected that the calibrated model fails to reproduce closely the reality and the capability to isolate faults will decrease. The situation can be mitigated by considering residuals between measurements and the estimated model responses or incorporating these residuals in the calibration process. The analysis of these possible scenarios and the verification of the real potential of the proposed solution in a more realistic setting is subject of further research.

\section*{Acknowledgements} \label{sec:acknowledgements}

This research was funded by the Swiss National Science Foundation (SNSF) Grant no. PP00P2 176878.

\bibliographystyle{unsrt}  
\bibliography{references}  


\makenomenclature
\nomenclature{$\text{HPC}$}{high-pressure compressor}
\nomenclature{$\text{HPT}$}{high-pressure turbine}
\nomenclature{$\text{LPC}$}{Low-pressure compressor}
\nomenclature{$\text{LPT}$}{Low-pressure turbine}
\nomenclature{$\text{N1}$}{Physical core speed}
\nomenclature{$\text{N2}$}{Physical fan speed}
\printnomenclature

\begin{figure}[ht]
\centering
\includegraphics[width=8.cm]{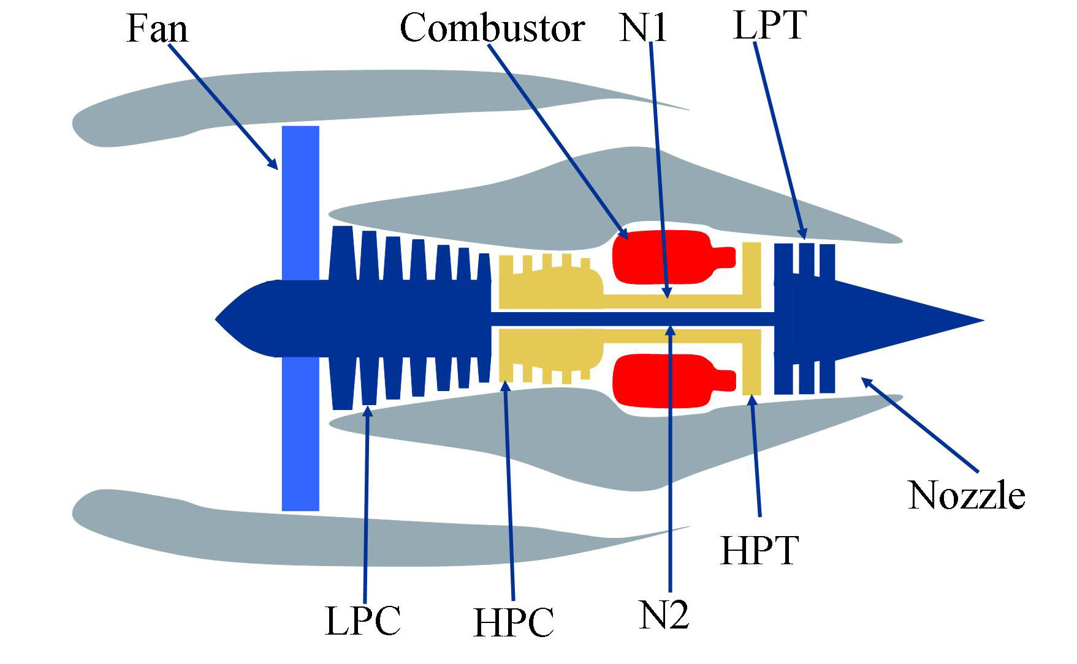}
\caption{Simplified diagram of the turbo fan engine model in C-MAPSS with Fan, Combustor, High-pressure compressor (HPC); high-pressure turbine (HPT); Low-pressure compressor (LPC); Low-pressure turbine (LPT); Physical core speed (N1), Physical fan speed (N2) and the Nozzle}
\label{fig:engine}
\end{figure}

\section{Appendix I: Neural Network Overview}
\label{sec:AppendixII}
In this section, we briefly introduce the selected discriminative and generative neural networks considered in our experiments. We focus first on discriminative models that try to learn $p(h_s|x)$ directly. In other words, algorithms that try to learn direct mappings from the space of inputs $X$ or $z$ to a label class (i.e. $\mathbf{T}$). In this group we introduce deep feed-forward networks (FF), vanilla autonecoders (AE) and hierarchical extreme learning machines (HELM). Finally, we will focus on generative algorithms that instead try to model the underlying distribution of the data $P(X)$ and show how these models can be combined with discriminative models to perform diagnostics tasks. In particular we introduce variational autoencoders (VAE).

\subsection{Discriminative models}
\textbf{Feed-forward neural network (FF)}. A deep feed-forward (FF) neural network with $L$ layers is a directed acyclic graph that implements a map $F: R^n \longmapsto R^{m^{L}}$ with the following structure:
\begin{align} \label{eq:ff}
   F &= F^{L} \circ F^{L-1} \circ  \dots \circ F^{1} \\
   F^{l} &= \sigma^{l} \circ  \overline{F}^{l} \\
   \overline{F}^{l}(x) &=  W^{l}x + b^{l} \in R^{m^{l}}
\end{align}
Hence, a feed-forward neural network represents a family of functions $F_{\mathcal{H}}$ parameterized by parameters $\mathcal{H} = \{{W^{l}, b^{l}}\}_{l=1}^{L}$ (i.e. weight matrices $W^l$ and biases $b^l$ for each layer). $\sigma_{l}$ denotes non-linear activation functions (e.g. tanh, ReLu, etc) and $\overline{F}^{l}$ denotes linear pre-activations. The number of neurons in each layer is given by $m^{l}$. We find the most appropriate function ($F_{\mathcal{H}}$) with the backpropagation algorithm \cite{LeCun2012} given a training set $S_T =\{x^{(i)}, y^{(i)}\}_{i=1}^N$ of $N$ input-output pairs.

The empirical risk on the training set $S_T$ is generally selected as optimisation metric for generation of discriminative models. The empirical risk minimizer is defined as: 
\begin{align}
   F_{\hat{\mathcal{H}}}(\mathcal{S}_{T}) = \arg\min_{\mathcal{H}} J(F_{\mathcal{H}}(x);\mathcal{S}_{T})
\end{align}
where $\hat{\mathcal{H}}$ corresponds to the optimal weights and bias of the neural network $F$ and $J(F;S_T)$ denotes the training risk of $F$ on the training sample $S_T$ 
\begin{align}
   J(F;S_T) = \frac{1}{N} \sum_{i=1}^N \ell(y^{(i)}, F(x^{(i)}))
\end{align}
\begin{equation}
	\ell(y,F(x)) =  \frac{1}{2}||y - F(x)||_2^2
\end{equation}

and the output target label $y$ corresponds to: $y=\mathbf{T}$ for the one-class network.

\textbf{Autonencoders (AE)}.
An autoencoder is any neural network that aims to learn the identity map (i.e. it is trained to reconstruct its own input). Therefore, it is a special case of the previous networks consisting of two parts with symmetric topology: an encoder ($E$) and a decoder ($D$). The encoder provides an alternative representation of the input ($x$) that we denote as $z$ and the decoder reconstructs back the input (i.e. $\overline{x}$) as closely as possible from its encoded representation $z$. The resulting mapping corresponds to the following structure:
\begin{align}
   F &= D \circ E\\
   E &= F^{l_z} \circ \dots \circ F^1 : \mathbb{R}^n \xrightarrow{} \mathbb{R}^d, x \xrightarrow{}z:x^l\\
   D &= F^L \circ \dots \circ F^{l_z+1}: \mathbb{R}^d \xrightarrow{} \mathbb{R}^n, z \xrightarrow{}\overline{x}
\end{align}

where the layer $l_z$ is generally a bottleneck (i.e. $d<n$) and therefore $z$ is a compressed representation of the input. Autoencoders can lean powerful non-linear generalization of principal component analysis (PCA).

The loss function of autoencoders is 
\begin{align}
   \ell(x, \overline{x}) &= \frac{1}{2}||x-F(x)||_2^2
\end{align}

\textbf{Hierarchical Extreme Learning Machines (HELM)}. Hierarchical Extreme Learning Machines are another popular neural network class for diagnostics task. Several researches have shown that it outperforms traditional machine learning method such us PCA and SVM in diagnostics task \cite{Michau}. HELM networks share similarities to three methods described earlier but with different topology and training method. As in deep RNN and FF networks, a HELM of $L$ layers has a hierarchy representations levels at each layer (i.e. $s^{l}$. This hierarchical hidden state $s^{l}$ that evolves as a function of the previous state $s^{l-1}$ defining a directed acyclic graph. However, in this case it evolves as a linear transformation.
\begin{align}
   s^{l} & = G^{l}(s^{l-1}, \beta^{l})  \in R^{\; m^{l}} \\
   G^{l}(s^{l-1}, \beta^{l}) &=  s^{l-1} {\beta^{l}}^{T} \quad  l=1, \dots, L-1
\end{align} 

with $s^0 := x$ 

The output of a HELM network is connected to the state of the last hidden layer $s^{L-1}$ as follows:
\begin{align} \label{eq:helm}
   F &= F^{L} \beta^{L} \\
   F^{L} & = \sigma^{L} \circ \overline{F}^{L}\\
   \overline{F}^{L} &=  W^{L}s^{L-1} +  b^{L} \\
   s^{L-1} &= G^{L-1} \circ  \dots \circ G^{1}(x)
\end{align}
Contrarily to previous networks the parameters $\mathcal{H} = \{{W^{l}, b^{l}}\}_{l=1}^{L}$ (i.e. weight matrices $W^l$ and biases $b^l$ for each layer) are random and are not optimised. Therefore, they provide an alternative (random) representation of the state $s^{l-1}$ (i.e. $F^{l}$) given weights $\{W^l, b^{l}\}$ and the non-linear transformation $\sigma^{l}$. The weight matrix $\beta^l$ are optimised layer wise to reconstruct the state $s^{l-1}$ from this random projection. Therefore, the loss function of $\beta$ resembles the auto encoder loss. However, typical regularization schemes are required correspond to the Maximum at Posterior (MAP)   
\begin{align}
   \beta^{l} &= \arg \min_{\beta^{l}} \lambda ||\beta^{l}||_{1} + ||F^{l}\beta^{l} - s^{l-1}||_{2}^{2} \\ 
   F^{l} &= \sigma^{l} \circ  \overline{F}^{l} \\
   \overline{F}^{l} &=  W^{l}s^{l-1} + b^{l}
\end{align}
with $s^L := y$

HELM are typically referred as autoencoder network due to the training process of the network, where the weight matrix $\beta$ is obtain from solving an autoencoder network for each of the hidden layers of HElM.

\subsection{Generative models}
Contrarily to the discriminate models that try to learn $p(h_s|x)$ directly, generative algorithms model the underlying distribution of the data $p(x)$. Concretely, generative latent models assume that an observed variable $x$ is generated by some random process involving an unobserved random (i.e. latent) variable $z$ \cite{Sarkar}. Hence, latent models define a joint distribution $p(x,z)=p(x|z)p(z)$ between a feature space $z$, and the input space $x$ \cite{Zhao}. Hence, the underling generation process resort to two steps: 1) a value $z^{(i)}$ is generated from some prior distribution $p(z)$ and 2) a value $x^{(i)}$ is generated from some conditional distribution $p(x|z)$. Hence, the data generation process is modeled with a complex conditional distribution $p_{\theta}(x|z)$, which is often parameterized with a neural network. There are two big families of generative models: generative adversarial networks (GANs) and Variational Autoencodes (VAEs). Our proposed method is based on VAEs that we explained in the following.

\textbf{Variational Autoencoder (VAE)}. Variational autoencoders \cite{Kingma2014} aim to sample values of $z$ that are likely to have produced $x$ and compute $p(x)$ from those \cite{Doersch2016}. As in the case of the standard vanilla autonecodes, VAE models comprise of an inference network (or encoder) and a generative network (or decoder). Contrarily to previews models, the latent representation $z$ of the data $x$ is a stochastic variable. Therefore, the encoder and the decoder networks are probabilistic. The inference network $q_{\phi} (z \vert x)$, parametrizes the intractable posterior $p(z \vert x)$ and the generative network $p_{\theta} (x \vert z)$ parametrizes the likelihood $p(x \vert z)$ with parameters $\theta$ and $\phi$ respectively. These parameters are the weights and biases of the neural network. A simple prior distribution $p(z)$ over the features is generally assumed (such us Gaussian or uniform).

The natural training objective of a generative model is to maximize the marginal likelihood of the data
\begin{align}
   \mathbb{E}_{p(x)}[\log p_{\theta}(x)] =  \mathbb{E}_{p(x)}[\mathbb{E}_{p(z)}[\log p_{\theta}(x \vert z)] 
\end{align}
However, direct optimization of the likelihood is intractable since $p_{\theta}(x) = \int_z p_{\theta}(x|z)p(z)dz$ requires integration \cite{Zhao}. Therefore, VAE consider the an approximation to the marginal likelihood denoted Evidence Lower BOund or ELBO; which is a lower bound to the log likelihood
\begin{align}
   \mathcal{L}_{\text{ELBO}} &= \mathbb{E}_{p(x)}[\mathbb{E}_{q_{\phi}(x|z)}[\log p_{\theta}(x \vert z)] - D_{KL}(q_\phi(z \vert x) \vert \vert p(z))]\\
   & \leq \mathbb{E}_{p(x)}[\log p_{\theta}(x)]
\end{align}
where $D_{KL}$ denotes the Kullback-Leibler. Hence, the training objective of VAE is to optimize the lower bound with respect to the variational parameters $\phi$ and the generative parameters $\theta$ 
\begin{align} \label{eq:ELBO}
   \max_{\phi, \theta} \mathbb{E}_{p(x)}[\mathbb{E}_{q_{\phi}(x|z)}[\log p_{\theta}(x \vert z)] - D_{KL}(q_\phi(z \vert x) \vert \vert p(z))]
\end{align}
The ELBO objective can be viewed as the sum of two components. The first term is the expected negative reconstruction error and it is similar to the training objective of a vanilla autoencoder. The $\text{KL}$ divergence ($D_KL \geq 0)$ is a distance measure of two probability distributions and acts as a regularizer of $\phi$ trying to keep the approximate posterior $q_{\phi}(z|x)$ close to the prior $p(z)$.  

\begin{figure}[ht]
\centering
\includegraphics[width=8cm]{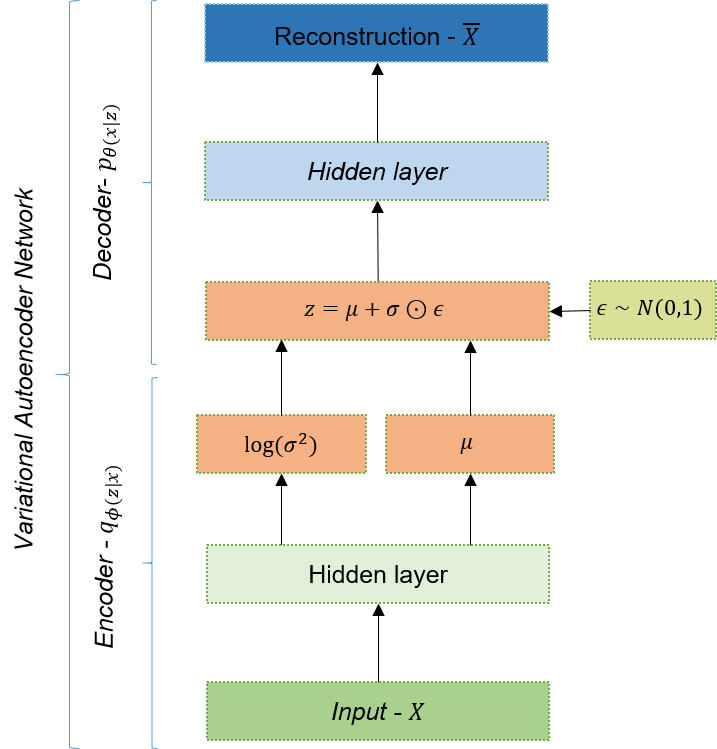}
\caption{Variational autoencoder network}
\label{fig:vae}
\end{figure}

Under certain hypothesis on the distribution families the KL divergence can be integrated analytically and therefore only the expected reconstruction error requires estimation by sampling. Therefore, direct optimization of $\mathcal{L}_{\text{ELBO}}$ with the back-propagation algorithm requires a good estimate of the gradient of the expectation $\nabla_{\phi}\mathbb{E}_{q_{\phi}(x|z)}[\log p_{\phi}(x \vert z)]$. However, naive Monte Carlo estimators exhibit very large variances and are therefore impractical. To find a low-variance gradient estimator a reparametrization of $z$ with a differentiable transformation $z = g(\epsilon,x)$ of an auxiliary noise variable $\epsilon$ is introduced \cite{Kingma2014}. The function $g(x, \epsilon)$  is generally chosen that maps an input datapoint $x^{(i)}$ and noise vector $\epsilon$ to a sample from the approximate posterior. The sampled $z^{(i)}$ is then input to the function $\log p_{\theta}(x|z)$ providing probability mass of a data point under the generative model $p_{\theta}$. 
Figure \ref{fig:vae} shows the resulting network architecture.
As a default assumption in VAE, the variational approximate posterior $q_{\phi}(z|x)$ follows a mutivariate Gaussian with diagonal covariance (i.e $q_{\phi}(z|x)= \mathcal{N}(z;\mu,\sigma^2 \mathbf{I})$). This assumption arises from the hypothesis that the true but intractable posterior $p_{\theta}(z|x)$ takes also the shape of an approximate Gaussian form with diagonal covariance. The distributions parameters of the approximate posterior $\mu$ and $\log \sigma^2$ are the non-linear embedding of the input $x$ provided by the encoder network with variational parameters $\phi$. Hence, the encoder output is a paramentrization of a approximate posterior distributions. Under these assumptions a valid local reparametrization of $z$ that allows to sample from the assumed Gaussian approximate posterior (i.e. $z^{(i)} \sim q_{\phi}(z|x^{(i)})$) is
\begin{align}
  z^{(i)} & = \mu^{(i)} + \sigma^{(i)} \odot\epsilon
\end{align}
with $\epsilon \sim \mathcal{N}(0,\mathbf{I})$. 

Since in this model we assume that both $p_{\theta}(z)$ and $q_{\phi}(z|x)$ are Gaussian distribution and therefore the $D_{KL}(q_\phi(z \vert x)$ can be computed analytically
\begin{align}
      -D_{KL}(q_\phi(z \vert x) \vert \vert p(z)) = \sum_{j=1}^d (1-\log (\sigma_j^{(i)})^2 - (\mu_j^{(i)})^2 -(\sigma_j^{(i)})^2\big)
\end{align}

\end{document}